\documentclass[journal,twoside,web]{ieeecolor}
\usepackage{generic}
\usepackage{cite}
\usepackage{amsmath,amssymb,amsfonts}
\usepackage{algorithmic}
\usepackage{graphicx}
\usepackage{textcomp}
\def\BibTeX{{\rm B\kern-.05em{\sc i\kern-.025em b}\kern-.08em
    T\kern-.1667em\lower.7ex\hbox{E}\kern-.125emX}}
\usepackage{epstopdf}
\usepackage{colortbl}
\usepackage{diagbox}
\usepackage{mathtools}
\usepackage{empheq}

\usepackage{pifont}
\definecolor{lightgray}{rgb}{0.83, 0.83, 0.83}

\newtheorem{mypbm}{Problem}

\usepackage[linesnumbered,boxed,commentsnumbered,ruled,vlined,longend]{algorithm2e}
\SetAlgorithmName{Procedure}
\usepackage{comment}
\usepackage[colorlinks = true,
linkcolor = blue,
urlcolor  = blue,
citecolor = blue,
anchorcolor = blue]{hyperref}

\DeclareMathOperator{\sign}{sign}

\title{Improving Power Systems Controllability via Edge Centrality Measures}

\author{MirSaleh Bahavarnia$^\dagger$, \IEEEmembership{Member, IEEE}, Muhammad Nadeem$^\dagger$, \IEEEmembership{Member, IEEE} and Ahmad F. Taha$^{\dagger,\star}$, \IEEEmembership{Member, IEEE}
	\thanks{$^\dagger$The authors are with the Department of Civil and Environmental Engineering, Vanderbilt University, 2201 West End Avenue, Nashville, TN 37235, USA. $^\star$Ahmad F. Taha is also affiliated with the Department of Electrical and Computer Engineering. Email addresses: mirsaleh.bahavarnia@vanderbilt.edu, muhammad.nadeem@vanderbilt.edu, ahmad.taha@vanderbilt.edu.}
 \thanks{This work is supported by the National Science Foundation, United States, under grants ECCS 2151571 and CMMI 2152450.}
}

\begin{document}

\maketitle

\begin{abstract}
Improving the controllability of power networks is crucial as they are highly complex networks operating in synchrony; even minor perturbations can cause desynchronization and instability. To that end, one needs to assess the criticality of key network components (buses and lines) in terms of their impact on system performance. Traditional methods to identify the key nodes/edges in power networks often rely on static centrality measures based on the network's topological structure ignoring the network's dynamic behavior. In this paper, using multimachine power network models and a new control-theoretic edge centrality matrix (ECM) approach, we: \textit{(i)} quantify the influence of edges (i.e., the line susceptances) in terms of controllability performance metrics, \textit{(ii)} identify the most \textit{influential} lines, and \textit{(iii)} compute near-optimal edge modifications that improve the power network controllability.   Employing various IEEE power network benchmarks, we validate the effectiveness of the ECM-based algorithm and demonstrate improvements in system reachability, control, and damping performance. 
\end{abstract}


\section{Introduction and Paper Contributions}\label{sec:Intro}

\IEEEPARstart{P}{ower} systems are vast and complex networks often represented as graph models comprising thousands of nodes (system buses) and edges (transmission lines). These nodes are classified into generators, loads, and non-unit nodes. Due to their susceptibility to fluctuations from complex load demands and the integration of renewable energy sources, enhancing the performance of power networks against various perturbations is crucial. One effective method to achieve this enhancement is through edge modifications. In power systems, edge modifications can be interpreted as changes in the impedance of transmission lines connecting the generators. This can be accomplished using flexible alternating current transmission system (FACTS) devices, such as thyristor-controlled series compensators \cite{zhang2012flexible}. 

Various \textit{centrality measures} have been proposed to identify the network's most central components (i.e., nodes or edges) in terms of performance variations against the perturbations. In most of the literature, the utilized centrality measures are mostly static as they solely consider the graph-theoretic topology (structure) of power networks, thereby overlooking the dynamic nature of power networks. Although significant advancements have been made in networked control theory in recent decades to quantify the importance of nodes/edges via centrality measures, some of this literature has not yet been investigated for power networks. In contrast, this paper investigates \textit{controllability improvement} of power networks via a newly proposed control-theoretic edge centrality measure. This specific problem has not been studied in power systems literature but there have been some related works. In the sequel, we review relevant works built upon centrality measures in both \textit{power networks} and \textit{control systems} literature.

In the power systems literature, various node/edge centrality measures have been proposed to analyze the structural vulnerability of the power networks (e.g., identifying the critical transmission lines) against perturbations like natural disasters or cyber-physical attacks. However, they have mostly been proposed based on the graph-theoretic definitions of node/edge\footnote{In this paper, adopting a Kron-reduction approach for extracting the susceptance matrix, by \textit{node} and \textit{edge} of the underlying graph of the power network, we mean the generator and susceptance value between paired generators, respectively \cite{wang2010electrical}.} centrality. Among them, we can list the following node/edge centrality measures: \textit{degree centrality} \cite{wang2010electrical,zio2010randomized,ernster2012power,caro2015centrality,chopade2016new}, \textit{eigenvector centrality} \cite{wang2010electrical,ernster2012power,caro2015centrality}, \textit{closeness centrality} \cite{wang2010electrical,nasiruzzaman2011application,ernster2012power,caro2015centrality}, and \textit{node/edge betweenness centrality} \cite{hines2008centrality,wang2010electrical,zio2010randomized,jin2010novel,nasiruzzaman2011application,ernster2012power,caro2015centrality,fang2018power,biswas2020graph}. 

Degree centrality ranks the nodes based on their degree (connectivity of the node to the rest of the network) in the graph representation of the network. Eigenvector centrality captures the importance of nodes in the network based on its adjacency matrix. Built upon a mean geodesic distance (i.e., the shortest path length in hops), closeness centrality determines how close a node is to the rest of the network. Node betweenness centrality specifies the mean fraction of the shortest paths between pairs of other nodes that pass through a node. In the aforementioned power network vulnerability analyses, the most influential system buses or transmission lines are ranked/sorted based on the corresponding values of the centrality measures so that the human operator can decide what actions on which components of the power network should be taken to mitigate the perturbations effectively. The common feature of all these graph-theoretic centrality measures is that they only take advantage of the graph-theoretic topology (structure) of power networks, thereby mainly overlooking the dynamic nature of power networks.

As for the recent control engineering literature, diverse edge centrality measures have been proposed to assess the importance of the edges/nodes in terms of their impacts on the performance metrics \cite{siami2017centrality,ghaedsharaf2021centrality,chanekar2021energy,chanekar2022encoding,chanekar2023gramian}. A significant advantage of such control-theoretic centrality measures over their graph-theoretic counterparts \cite{freeman1977set,bonacich1987power,brandes2001faster}, is that in addition to the network topology (structure), they also consider the effects of the dynamics among the agents on the centrality of the edges/nodes. Specifically, for the continuous-time linear consensus networks (without and with time delays, respectively), \cite{siami2017centrality,ghaedsharaf2021centrality} elaborate on proposing an $\mathcal{H}_2$ norm-based approach to quantify the impact of each edge on the network's $\mathcal{H}_2$ performance. In \cite{chanekar2022encoding}, considering the discrete-time linear networks (not necessarily consensus networks), the authors delve into encoding edge perturbation impact on controllability via \textit{edge centrality matrix} (ECM) approach. We highlight that none of the aforementioned control-theoretic edge centrality measures have been investigated or applied to the power systems yet. Nevertheless, in the case of node centrality measures, \cite{chanekar2023gramian} has examined the effectiveness of a controllability-Gramian-based node centrality measure, namely \textit{vulnerability matrix (VM)}, in terms of vulnerability assessment against the nodal impulse inputs affecting flows of a power network.

ECM approach \cite{chanekar2022encoding} has been introduced to facilitate the computation of the perturbation impact of each edge on the network controllability. The authors in \cite{chanekar2022encoding} introduce a novel systematic way to construct ECM for discrete-time systems with linear dynamics and propose computationally efficient alternatives to compute ECM for various controllability-Gramian-based performance metrics including the trace, log-det, and negated trace inverse. They also derive theoretical bounds on the controllability-Gramian-based performance metrics and collect empirical observations for some special class of networks including stem-bud networks and Erd\H{o}s-Rényi networks. Through various numerical experiments, they show that the ECM-based edge modification method has a near-optimal performance while significantly improving the computational time. In \cite{chanekar2021energy}, considering the adjacency-type networks, the authors propose theoretical results on the optimal solutions under some assumptions on the controllability Gramian and type of the chosen controllability-Gramian-based performance metric. Nevertheless, no performance guarantees are provided due to the complexity of the optimization problem. These purely control-theoretic methodologies motivate us to take advantage of ECM and employ it to analyze the vulnerability of power networks against perturbations and effectively enhance the robustness and efficiency of power networks.

A similar line of research exists in power network stability analysis that elaborates on quantifying the sensitivity (mainly via differential calculus) of power network stability to the perturbations \cite{overbye2013power,marzooghi2020scenario,brahma2020sensitivity,zeng2023stability}. The study \cite{overbye2013power} has presented a computationally efficient algorithm
for determining the set of most influential transmission lines that contribute to the effective geomagnetically induced currents in a given set of transformers. The research work \cite{marzooghi2020scenario} utilizes a novel scenario-sensitivity-contingency-based framework to evaluate the system stability along possible evolution pathways toward highly renewable future power grids. In \cite{brahma2020sensitivity}, a quantitative dynamic
flexibility index is presented capturing the effect of varying bus injections on the system's small-signal stability. In \cite{zeng2023stability}, built upon the \textit{Greshgorin Theorem}, the sensitivity analysis is utilized for analyzing the influencing factors to improve system stability. As a common approach among the aforementioned research works, the authors first define and calculate a sensitivity measure and then rank the electrical parameters based on the associated sensitivity values. Afterward, they propose a sensitivity rank-based method to effectively mitigate the perturbations accordingly. Again, the current study differs from these sensitivity-based works as we take advantage of the dynamics of the network in addition to the topology (structure) of the network. Moreover, as another difference, we utilize an ECM approach specialized for power networks modeled by swing dynamics consisting of a Laplacian-like susceptance matrix.

\noindent \textbf{Paper Contributions.} In this paper, taking advantage of the swing dynamics of the synchronous generators and adopting the continuous-time version of a control-theoretic ECM approach utilized by \cite{chanekar2022encoding}, we contribute to the networked power system literature as follows:
\begin{itemize}
    \item We quantify the influence of edges (susceptance values between paired generators) of power networks via the controllability-Gramian-based performance metrics (including the trace, log-det, and negated trace inverse).
    \item The proposed approach first identifies the key edges/transmission lines that most affect the power network’s controllability (in terms of defined controllability Gramian metrics). Then, we determine the best possible modifications to these lines to enhance the network’s controllability. Remarkably, since identifying the optimal edge modification vector requires a brute-force exhaustive combinatorial search (the NP-hardness), the proposed ECM-based approach significantly reduces the computational burden.
    
    \item We validate the effectiveness of the ECM-based results by employing various IEEE power network benchmarks. We also evaluate the near-optimality via defined near-optimality performance measures in the case of computationally feasible circumstances (e.g., the low-cardinality edge modification scenario). Furthermore, we additionally assess the relative performance of the proposed ECM-based method compared to a static graph-theoretic edge centrality-based alternative, namely nearest neighbor edge centrality (NNEC) \cite{brohl2022straightforward}, in identifying the network's most influential edges.

\end{itemize}

\noindent \textbf{Paper Organization.} The remainder of the paper is structured as follows: Section \ref{sec:ProFor} presents the preliminaries on the edge perturbation impact and elaborates on an ECM-based perturbation impact assessment followed by the problem statement. Section \ref{ECMBRM} sheds light on an ECM-based edge modification design. Section \ref{secNS} empirically verifies the effectiveness of the proposed ECM-based perturbation impact assessment and edge modification design via conducting a comprehensive set of numerical experiments on various IEEE power network benchmarks. Finally, Section \ref{Con} provides a few concluding remarks along with some pertinent future directions.

\noindent {\bf Paper Notation.} We show the set of real numbers and the set of positive real numbers by $\mathbb{R}$ and $\mathbb{R}_{++}$, respectively. We denote the set of real-valued $n$-dimensional vectors and $m \times n$ matrices by $\mathbb{R}^n$ and $\mathbb{R}^{m \times n}$, respectively. We show the matrix/vector transpose by $^\top$. We denote the $j$-th canonical unit vector by $e_j$ and define the unit matrix $E_{ji}$ as $E_{ji} := e_j e_i^\top$. For a vector $v$, we represent the diagonal matrix associated with $v$ by $\mathbf{diag}(v)$. Given a square matrix $M = [m_{ji}]$, let $\alpha(M)$, $\mathrm{tr}(M)$, and $\log \det (M)$ respectively denote the spectral abscissa of $M$ (i.e., the maximum real part of the eigenvalues of $M$), the trace of $M$, and the logarithm of determinant of $M$. Moreover, if $M$ is invertible, $M^{-1}$ and $M^{-2}$ respectively denote the inverse of $M$ and the inverse-square of $M$. The notations $I_n$ and $\mathbf{1}_{n}$ represent the $n$-dimensional identity matrix and $n$-dimensional all-one vector, respectively. For a matrix $M$, let $|M|$ and $\|M\|$ respectively denote the element-wise absolute value of $M$ and the spectral norm of $M$. To show matrix element-wise product (i.e., Hadamard product), we utilize symbol $\odot$. We represent the positive definiteness and positive semi-definiteness by $\succ 0$ and $\succeq 0$, respectively. For a complex number $z$, let $\Re(z)$ and $\Im(z)$ show the real and imaginary parts, respectively. We use $\jmath$ to denote the imaginary unit $\sqrt{-1}$. For a given natural number $q$, let us denote $\{1,\dots,q\}$ by $\mathbb{N}_q$. For a finite set $S$, we denote its cardinality (number of the elements) by $|S|$. We show the element-wise vector inequality by standard notation $\ge$. We denote the computational complexity by $\mathcal{O}$. Symbols $\land$ and $\neg$ show the logical \texttt{AND} and logical \texttt{NOT} operations. We denote the set membership and subset-ness by $\in$ and $\subseteq$, respectively. To represent the set intersection and set difference, we respectively utilize notations $\cap$ and $\setminus$. We denote the expected value and normal distribution with zero mean and unit covariance with $\mathbf{E}(.)$ and $\mathcal{N}(0,I)$, respectively.

\section{Preliminaries and Problem Statement} \label{sec:ProFor}

In Section \ref{LSSM}, we detail a model of power system dynamics followed by a state-space representation and average mode elimination in Section \ref{SecSsr}. Some preliminaries on edge perturbation impact for power systems are provided in Section \ref{SFI}. Finally, built upon an ECM-based perturbation impact assessment presented by Section \ref{ECMBRA}, Section \ref{SPS} formally states the research problem.

\subsection{Power system dynamics} \label{LSSM}

Considering the second-order generator dynamics, and for constant-current and constant-impedance loads, the nonlinear differential-algebraic equations (NDAEs) of the power system dynamics can be simplified and reduced to the electromechanical swing dynamics of the synchronous generators \cite{dorfler2014sparsity,kundur1994power}. The swing dynamics for bus $j$ are given as follows:
{\begin{align} \label{SwEq}
    & M_{j} \ddot{\theta}_j(t) + D_{j} \dot{\theta}_j(t) + \sum_{i = 1}^{N} |y_{ji}|\mathbb{E}_j \mathbb{E}_i \sin (\theta_j(t)-\theta_i(t)-\varphi_{ji}) \notag\\
    & = P_{j}(t),
\end{align}}where $M_j$, $\theta_k(t)$, $D_j$, $N$, $Y = [y_{ji}]$, $\mathbb{E}_k$, $\varphi_{ji}$, and $P_j(t)$ denote the $j$-th inertia coefficient, $k$-th rotor angle, $j$-th damping coefficient, number of synchronous generators, Kron-reduced admittance matrix, $k$-th q-axis voltage, phase shifts associated with pair $(i,j)$, and $j$-th generator injection power, respectively. The phase shifts are related to the Kron-reduced admittance matrix via $\varphi_{ji} = -\sign(j-i) \arctan \Big(\frac{\Re(y_{ji})}{\Im(y_{ji})}\Big)$.

Linearizing \eqref{SwEq} around an equilibrium $(\theta^{(e)},\dot{\theta}^{(e)})$, one gets
\begin{align} \label{LiSwEq}
    M \ddot{\theta}(t) + D \dot{\theta}(t) + L \theta(t) = 0,
\end{align}
where $M \in \mathbb{R}^{N \times N}$, $D \in \mathbb{R}^{N \times N}$, and $L \in \mathbb{R}^{N \times N}$ denote the diagonal matrix of generator inertia coefficients, diagonal matrix of generator damping coefficients, and Laplacian matrix of susceptance values among the generators, respectively. Laplacian matrix of susceptance values among the generators $L$ must be symmetric because the underlying graph of the electrical network is undirected \cite{dorfler2012kron}. The off-diagonal and diagonal elements of $L \succeq 0$ are as follows:
\begin{align}
    l_{ji} &= \begin{cases}
       -|y_{ji}|\mathbb{E}_j\mathbb{E}_i \cos (\theta_j^{(e)}-\theta_i^{(e)}-\varphi_{ji}) & \mathrm{if}~j \neq i,\\
       -\sum_{i=1, i \neq k}^N l_{ki} & \mathrm{else}.
    \end{cases}
\end{align}

For any $j \neq i$, it is noteworthy that $l_{ji} = l_{ij}$ holds because $y_{ji} = y_{ij}$, $\varphi_{ji} = -\varphi_{ij}$, and $\cos (\theta_j^{(e)}-\theta_i^{(e)}-\varphi_{ji}) = \cos (\theta_i^{(e)}-\theta_j^{(e)}+\varphi_{ji})$ are satisfied. Note that in \cite{dorfler2014sparsity}, the authors have used $\varphi_{ji} = -\arctan \Big(\frac{\Re(y_{ji})}{\Im(y_{ji})}\Big)$ as a formula for the phase shift. Considering such a formula for the phase shift implies $\varphi_{ji} = \varphi_{ij}$ and can erroneously lead to an asymmetric $L$ (which is incorrect as $L$ must be a symmetric Laplacian matrix) because $\cos (\theta_j^{(e)}-\theta_i^{(e)}-\varphi_{ji}) = \cos (\theta_i^{(e)}-\theta_j^{(e)}-\varphi_{ji})$ does not necessarily hold. To resolve such an issue, we incorporated the term $\sign(j-i)$ to obtain $\varphi_{ji} = -\sign(j-i) \arctan \Big(\frac{\Re(y_{ji})}{\Im(y_{ji})}\Big)$.

The linear electromechanical swing dynamics of the synchronous generators \cite{dorfler2014sparsity,kundur1994power} have widely been utilized in the networked control of power networks. Such simple linear power dynamics (linear electromechanical swing dynamics) provide us with a convenient framework to formulate/solve optimal control problems with the susceptance values (edges) of the susceptance (Laplacian) matrix $L$ as the optimization variables. For instance, \cite{wu2014sparsity,wu2016input,bhela2021efficient} propose a sparsity-promoting $\mathcal{H}_2$ control to optimally modify the susceptance values subject to low/limited communication requirements. Utilizing diverse mathematical tools like the alternating direction method of multipliers (ADMM) and the mixed-integer linear programming (MILP), the authors of \cite{wu2014sparsity,wu2016input,bhela2021efficient} obtain near-optimal solutions to the original NP-hard problem of the optimal design of the network topology, respectively.

\subsection{State-space representation and average mode elimination} \label{SecSsr}

According to \eqref{LiSwEq} that is a set of second-order ordinary differential equations (ODEs), considering $\dot{\theta}(t) = \omega(t)$ with $\theta(t)$ as the rotor angles and $\omega(t)$ as the rotor frequencies, and additionally incorporating the corresponding terms for the fast electrical dynamics associated with the generator excitation via power system stabilizers (PSSs) and governor control, we can construct the following state-space representation of the linear power dynamics \cite{wu2014sparsity,kundur1994power}:
{\begin{subequations} \label{PNSS}
    \begin{align}
    & \dot{\phi}(t) = \bar{A}(L) \phi(t) + \bar{B}u(t),\\
    & \bar{A}(L) = \begin{bmatrix}
        0 & I_N & 0\\
        -M^{-1}L & -M^{-1}D & \bar{A}_{vr}\\
        \bar{A}_{rp} & \bar{A}_{rv} & \bar{A}_{rr}
    \end{bmatrix},~\bar{B} = \begin{bmatrix}
        0 \\
        M^{-1}\\
        0
    \end{bmatrix},
    \end{align}
\end{subequations}}where $\phi(t) = \begin{bmatrix}
    \theta(t)^\top & \omega(t)^\top & r(t)^\top
\end{bmatrix}^\top \in \mathbb{R}^n$ denotes the state vector consisting of the rotor angles $\theta(t) \in \mathbb{R}^N$, rotor frequencies $\omega(t) \in \mathbb{R}^N$ (note that $\dot{\theta}(t) = \omega(t)$ holds), and fast electrical dynamics $r(t) \in \mathbb{R}^{n-2N}$, and $u(t) \in \mathbb{R}^N$, $\bar{A}(L) \in \mathbb{R}^{n \times n}$, and $\bar{B} \in \mathbb{R}^{n \times N}$ represent the input vector, state matrix, and input matrix, respectively. The matrices $\bar{A}_{vr}$, $\bar{A}_{rp}$, $\bar{A}_{rv}$, and $\bar{A}_{rr}$ denote the matrices consisting of the coefficients associated with the fast electrical dynamics \cite{wu2014sparsity}.

Because $L$ is a Laplacian matrix, it can be represented as $L = \mathbf{diag}(G \mathbf{1}_N)-G$ where $G = [g_{ji}] \in \mathbb{R}^{N \times N}$ denotes the adjacency matrix associated with $L$. Since $G$ is a symmetric matrix, i.e., $G = G^\top$ holds, from now on, we consider the set of $g_{ji}$s with $j < i$ for which the cardinality is equal to $\frac{N(N-1)}{2}$. Note that $\bar{A}(L)$ has a $0$ eigenvalue, because $L \mathbf{1}_N = 0_N$ and $\bar{A}_{rp} \mathbf{1}_N = 0_{n-2N}$ are satisfied \cite{wu2014sparsity} and we subsequently have
{\begin{align*}
    & \begin{bmatrix}
        0 & I_N & 0\\
        -M^{-1}L & -M^{-1}D & \bar{A}_{vr}\\
        \bar{A}_{rp} & \bar{A}_{rv} & \bar{A}_{rr}
    \end{bmatrix} \begin{bmatrix}
        \mathbf{1}_N \\0_{N}\\ 0_{n-2N}
    \end{bmatrix} = 0.
\end{align*}}To eliminate the marginal stability of $\bar{A}(L)$ due to the $0$ eigenvalue, we need to eliminate the average mode accordingly. We see later on it helps us to obtain a well-defined controllability Gramian. 

As detailed by \cite{wu2014sparsity}, to eliminate the average mode, $\tilde{\theta}(t) = \frac{1}{N} \mathbf{1}_N^\top \theta(t)$, from \eqref{PNSS}, one can utilize the following coordinate transformation:
\begin{align} \label{phix}
	& \underbrace{\begin{bmatrix}
			\theta(t) \\ q(t)
	\end{bmatrix}}_{\phi(t)} = \underbrace{\begin{bmatrix}
			U & 0\\0 & I_{n-N}
	\end{bmatrix}}_{T} \underbrace{\begin{bmatrix}
			\psi(t) \\
			q(t)
	\end{bmatrix}}_{x(t)} + \begin{bmatrix}
		\mathbf{1}_N\\0_{n-N}
	\end{bmatrix}\tilde{\theta}(t),
\end{align}
where $q(t) = \begin{bmatrix}
	\omega(t)^\top & r(t)^\top
\end{bmatrix}^\top \in \mathbb{R}^{n-N}$, $U \in \mathbb{R}^{N \times (N-1)}$, $\psi(t) \in \mathbb{R}^{N-1}$, and $x(t) \in \mathbb{R}^{n-1}$. Particularly, $U$ can be construed by stacking the $N-1$ eigenvectors of $I_N - \frac{1}{N}\mathbf{1}\mathbf{1}^\top$ as the columns of $U$. The following identities
\begin{align*}
	U^\top U = I_{N-1},~U U^\top = I_N - \frac{1}{N}\mathbf{1}\mathbf{1}^\top,~U^\top \mathbf{1}_N = 0_{N-1},
\end{align*}
imply that $x(t) = T^\top \phi(t)$ holds. Substituting \eqref{phix} into \eqref{PNSS} and noting that $L \mathbf{1}_N = 0_N$ and $\bar{A}_{rp} \mathbf{1}_N = 0_{n-2N}$ are satisfied \cite{wu2014sparsity}, we obtain:
\begin{equation} \label{PNSSReduced}
\boxed{\dot{x}= A(L) x+ B u,~A(L) = T^\top \bar{A}(L) T,~B=T^\top \bar{B},}
\end{equation}where $A(L) \in \mathbb{R}^{(n-1) \times (n-1)}$ and $B \in \mathbb{R}^{(n-1) \times N}$.

\subsection{Edge perturbation impact for power systems} \label{SFI}

Consider the continuous-time LTI power system described by the state-space \eqref{PNSSReduced}. We assume that pair $(A(L),B)$ is controllable. It is noteworthy that $A(L)$ is stable (Hurwitz) if $\alpha(A(L)) < 0$ holds. As discussed earlier, to capture the edge perturbation impact, we utilize a control-theoretic ECM approach \cite{chanekar2022encoding}. To compute ECM, we need preliminaries on the controllability as detailed in the sequel.

The controllability of \eqref{PNSSReduced} is equivalent to the ability to steer the state from an initial state $x(0) = x_0$ to any arbitrary final state $x(t_f) = x_f$ in $t_f$ time unit. To verify the controllability of \eqref{PNSSReduced}, one may consider the following controllability Gramian \cite{chen1984linear}:
\begin{align} \label{CG}
    W_c(t_f) &:= \int_{0}^{t_f} e^{A(L) t} B B^\top e^{A(L)^\top t}~dt.
\end{align}
Pair $(A(L),B)$ is controllable if the controllability Gramian $W_c(t_f)$ in \eqref{CG} is positive definite, i.e., $W_c(t_f) \succ 0$ holds.

To capture the quantitative nature of the controllability, one can employ three key controllability performance metrics \cite{pasqualetti2014controllability,chanekar2022encoding}:{\begin{subequations} \label{PerfoMet}
\begin{align}
    & \textrm{trace:}~\mathrm{tr}(W_c(t_f)),\\
    & \textrm{log-det:}~\log \det(W_c(t_f)),\\
    & \textrm{negated trace inverse:}~-\mathrm{tr}(W_c(t_f)^{-1}).
\end{align}     
\end{subequations}}All these performance metrics have energy-based interpretations built upon the fact that the minimum energy required to steer the state from $x_0$ to $x_f = 0$ in $t_f$ is equal to $x_0^\top W_c(t_f)^{-1} x_0$ \cite{pasqualetti2014controllability,chanekar2022encoding}. Then, the higher the value of any of the performance metrics, the less minimum energy required to steer the state from $x_0$ to $x_f = 0$. Such an observation suggests the maximization of the performance metrics to significantly improve the controllability of the system. The \textit{edge perturbation impact} matrix $\mathcal{I} = [\mathcal{I}_{ji}]$ is defined as
\begin{align} \label{FIM}
    \mathcal{I} &:= |\Upsilon|,
\end{align}
where $\Upsilon = [\upsilon_{ji}]$ denotes ECM \cite{chanekar2022encoding}. The higher value $\mathcal{I}_{ji}$ we have, the more influential edge $(i,j)$ we have. Later on, we elaborate on how to construct ECM $\Upsilon$ which is built upon the first-order perturbation analysis to measure the first-order impact of edge perturbation on network controllability performance \cite{chanekar2022encoding}. 

\subsection{An ECM-based perturbation impact assessment} \label{ECMBRA}

In this section, to quantitatively assess the ECM-based perturbation impact \eqref{FIM}, we construct ECM $\Upsilon$ specialized for \eqref{PNSSReduced} along with various controllability-Gramian-based performance metrics emphasizing the fact that the edges of susceptance (Laplacian) matrix $L$ are subject to perturbation. In the sequel, we first compute the gradient of the performance metrics by computing the controllability Gramian and its partial derivatives with respect to the edges. We then construct ECM and the ECM-based perturbation impact matrix.

Here, we elaborate on how to construct ECM $\Upsilon$ for power system dynamics modeled by \eqref{PNSSReduced}. To assess the edge perturbation impact, we compute the gradient of performance metrics with respect to the susceptance-dependent parameters $g_{ji}$ with $g_{ji} = -l_{ji}$ for $j < i$ (i.e., first-order perturbation analysis). For a controllability Gramian $W_c(t_f)$, one may compute the gradient of performance metrics in \eqref{PerfoMet} as follows \cite{chanekar2022encoding}:{\begin{subequations} \label{PMsn}
\begin{align}
    \frac{\partial}{\partial g_{ji}} \mathrm{tr}(W_c(t_f)) &= \mathrm{tr}\bigg(\frac{\partial W_c(t_f)}{\partial g_{ji}}\bigg),\\
    \frac{\partial}{\partial g_{ji}} \log \det(W_c(t_f)) &= \mathrm{tr}\bigg(W_c(t_f)^{-1} \frac{\partial W_c(t_f)}{\partial g_{ji}}\bigg),\\
    \frac{\partial}{\partial g_{ji}} (-\mathrm{tr}(W_c(t_f)^{-1})) &= \mathrm{tr}\bigg(W_c(t_f)^{-2} \frac{\partial W_c(t_f)}{\partial g_{ji}} \bigg).
\end{align}
\end{subequations}}Considering \eqref{PNSSReduced} and noting that $A(L)$ is stable (Hurwitz), we can obtain the controllability Gramian $W_c(t_f)$ in \eqref{CG} for $t_f = \infty$ (simply denoting $W_c(t_f = \infty)$ with $W_c$) via solving the following continuous-time Lyapunov equation:
\begin{align} \label{RLEn}
    & A(L)W_c + W_c A(L)^\top + B B^\top = 0.
\end{align}
For $j < i$, taking the partial derivative from \eqref{RLEn} with respect to $g_{ji}$ and defining notations $X_{ji} := \frac{\partial W_c}{\partial g_{ji}}$ and $F_{ji} := \frac{\partial A(L)}{\partial g_{ji}}$, we get{\begin{subequations} \label{WGn}
    \begin{align} 
    & A(L)X_{ji} + X_{ji}A(L)^\top + F_{ji}W_c + W_cF_{ji}^\top = 0, \label{WG1n}\\
    & F_{ji} = T^\top \frac{\partial \bar{A}(L)}{\partial g_{ji}} T = T^\top \begin{bmatrix}
        0 & 0 & 0\\
        -M^{-1}V_{ji} & 0 & 0\\
        0 & 0 & 0
    \end{bmatrix}T, \label{WG2n}\\
    & V_{ji} := E_{jj} - E_{ji} - E_{ij} + E_{ii},~(E_{ji} := e_j e_i^\top). \notag
\end{align}
\end{subequations}}Solving the Lyapunov equation \eqref{RLEn} for $W_c$, one can first evaluate $F_{ji}W_c + W_cF_{ji}^\top$ and then solve \eqref{WGn} for $X_{ji}$.

We now define $\mathcal{S}_N := \{(i,j) \in \mathbb{N}_N \times \mathbb{N}_N:i > j\}$. We impose $i > j$ as the graph of the power network is an undirected graph. Computing $X_{ji}$ via solving the Lyapunov equation \eqref{WGn} and utilizing \eqref{PMsn}, one may respectively construct ECM $\Upsilon = [\upsilon_{ji}]$ associated with the three performance metrics $\mathrm{tr}(W_c)$, $\log \det(W_c)$, and $-\mathrm{tr}(W_c^{-1})$ as
\begin{subequations} \label{ECMCn}
    \begin{align}
        \upsilon_{ji} &:= \mathrm{tr}(X_{ji}),~\forall(i,j) \in \mathcal{S}_N,\\
        \upsilon_{ji} &:= \mathrm{tr}(W_c^{-1} X_{ji}),~\forall(i,j) \in \mathcal{S}_N,\\
        \upsilon_{ji} &:= \mathrm{tr}(W_c^{-2} X_{ji}),~\forall(i,j) \in \mathcal{S}_N.
    \end{align}
\end{subequations}To portray the importance of edge $(i,j)$ in terms of its impact on the aforementioned controllability-Gramian-based performance metrics in \eqref{PerfoMet}, we first compute the edge perturbation impact matrix $\mathcal{I}$ via \eqref{FIM}, i.e., $\mathcal{I} := |\Upsilon|$. Given a candidate edge set, namely $\mathcal{E} \subseteq \mathcal{S}_N$, by forming the sorted sequence (with a descending order) of $\mathcal{I}_{ji} = |\upsilon_{ji}|$ for all $(i,j) \in \mathcal{E}$, namely $\{\tau_l\}_{l=1}^{|\mathcal{E}|}$ with $\tau_1 \ge \dots \ge \tau_{|\mathcal{E}|}$, we can realize the importance of edge $(i,j)$ in terms of its impact on the corresponding controllability-Gramian-based performance metrics in \eqref{PerfoMet}. The philosophy behind defining a candidate edge set and potentially choosing a proper subset of edges could be the fact that in some circumstances, we are unable to modify some edges due to some physics-induced limitations. It is noteworthy that the cardinality of such a sorted sequence is less than or equal to $\frac{N(N-1)}{2}$. A common way to select a candidate edge set $\mathcal{E}$ is choosing the edges with non-zero susceptance values $l_{ji}$s with $i > j$. Such a candidate edge set $\mathcal{E}$ is the edge set associated with $L$, namely $\mathcal{E}^L$.

    In the case of $\mathcal{E} = \mathcal{S}_N$, the computational complexity of computing ECM $\Upsilon$ is $\mathcal{O}(N^5)$ as it requires $\mathcal{O}(N^2)$ operations of solving the Lyapunov equation \eqref{WGn} (which is of $\mathcal{O}(N^3)$ computational complexity per each edge $(i,j)$). However, for $\mathcal{E} \subset \mathcal{S}_N$, the computational complexity of computing ECM $\Upsilon$ is $\mathcal{O}(|\mathcal{E}|N^3)$ as for all $(i,j) \notin \mathcal{E}$ one can set $\upsilon_{ji} = 0$.

\subsection{Problem statement} \label{SPS}

Built upon the definition of edge perturbation impact matrix $\mathcal{I}$ in \eqref{FIM} and considering the ECM $\Upsilon = [\upsilon_{ji}]$ formulas presented by \eqref{ECMCn}, we formally state the following problem for the power system dynamics \eqref{PNSSReduced}:
\begin{mypbm} \label{Prob1}
    Given the power system dynamics \eqref{PNSSReduced} and considering the ECM $\Upsilon = [\upsilon_{ji}]$ formulas presented by \eqref{ECMCn}, compute the ECM-based perturbation impact \eqref{FIM} and design edge modifications to improve system controllability.
\end{mypbm} 

To improve the controllability of the power network stated by Problem \ref{Prob1}, we utilize nonlinear optimization tools to solve for a near-optimal edge modification vector aiming at improving the controllability of the power network. In the next section, we construct an ECM-based maximization problem and solve it for a near-optimal edge modification vector. For brevity, we utilize a unified notation $h(W_c(A(\mathcal{L})))$ for all the controllability-Gramian-based performance metrics in \eqref{PerfoMet} where $W_c(A(\mathcal{L}))$ denotes the $\mathcal{L}$-dependent solution of the continuous-time Lyapunov equation \eqref{RLEn} replacing $A(L) = T^\top \bar{A}(L) T$ with $A(\mathcal{L}) = T^\top \bar{A}(\mathcal{L})T$. Also, we consider the auxiliary notations/definitions associated with the edge modification summarized in Tab. \ref{tab:my_labelS}.

\begin{table}[t]
    \centering
    \caption{Auxiliary notations/definitions associated with the edge modification.}
    {\begin{tabular}{|c|c|}
       \hline
        $\mathcal{E}$ & $\mathrm{Candidate~edge~set},~\mathcal{E} \subseteq \mathcal{S}_N$\\
        \hline
        $\mathcal{E}^{L}$ & $\mathrm{Edge~set~associated~with}~L,~\mathcal{E}^{L} \subseteq \mathcal{S}_N$\\
        \hline
        $s$ & $\mathrm{Number~of~modified~edges}$ \\
        \hline
        $\mathcal{E}_s$ & $\mathrm{Edge~modification~set},~\mathcal{E}_s \subseteq \mathcal{E},~|\mathcal{E}_s|=s$\\
        \hline
        $\gamma$ & $s$-$\mathrm{dimensional~edge~modification~vector}$\\
        \hline
        $\gamma_{k_{ji}}$ & $\mathrm{Element~of~} \gamma \in \mathbb{R}^{s} \mathrm{~associated~with~edge~} (i,j)$\\
        \hline
        $\Delta(\gamma)$ & $\mathrm{Edge~modification~matrix}$: $\sum_{(i,j) \in \mathcal{E}_s} \gamma_{k_{ji}} V_{ji}$\\
        \hline
    \end{tabular}}
    \label{tab:my_labelS}
\end{table}

\section{An ECM-based Edge Modification Design} \label{ECMBRM} 

\subsection{Edge modification design formulation} \label{EMD1}

Given $s \in \mathbb{N}_{|\mathcal{E}|}$, $\beta \in \mathbb{R}_{++}$, and an edge modification set $\mathcal{E}_s$, we consider the following optimization problem:{\begin{subequations} \label{OPn}
   \begin{align}
    \underset{\gamma \in \mathbb{R}^{s}}{\max}&~h(W_c(A(L + \Delta(\gamma)))), \label{OPnobj}\\
\mathrm{subject~to}&~\alpha(A(L + \Delta(\gamma))) < 0,\label{StabCon}\\
&~\|\gamma\| \le \beta, \label{OPCn}\\
&~\gamma \ge \iota_{\mathcal{E}_s}(L), \label{iotaL} 
    \end{align} 
\end{subequations}}where the $k_{ji}$-th element of $\iota_{\mathcal{E}_s}(L) \in \mathbb{R}^s$ is set as $l_{ji} = -g_{ji}$ for any $(i,j) \in \mathcal{E}_s$. In optimization problem \eqref{OPn}, the objective function in \eqref{OPnobj} reflects the controllability-Gramian-based performance metric to be improved. The inequality constraint \eqref{StabCon} is the stability constraint ensuring the stability of $A(L + \Delta(\gamma))$ and subsequently well-posedness of $W_c(A(L + \Delta(\gamma)))$ in the objective function in \eqref{OPnobj}. The inequality constraint \eqref{OPCn} can be interpreted as a bounded budget/energy of the edge modification vector $\gamma$. The inequality constraint \eqref{iotaL} is enforced to guarantee the non-negativity of the modified edges, i.e., $\gamma_{k_{ji}} + g_{ji} \ge 0$ for any $(i,j) \in \mathcal{E}_s$. It is noteworthy that $\Delta(\gamma)$ can also be represented as $\Delta(\gamma) = C \mathbf{diag}(\gamma)C^\top$ where $C \in \mathbb{R}^{N \times s}$ denotes the incidence matrix associated with the edge modification set $\mathcal{E}_s$ defined via the following rules for $(i,j) \in \mathcal{E}_s$: $C_{jk_{ji}} = 1$, $C_{ik_{ji}} = -1$, and $C_{lk_{ji}} = 0$ for $(l \neq j) \land (l \neq i)$.

We highlight that since edge modification matrix $\Delta(\gamma)$ appearing in the objective function \eqref{OPnobj} and the stability constraint \eqref{StabCon} depends on edge modification set $\mathcal{E}_s$ and additional bounding constraints \eqref{OPCn} and \eqref{iotaL} must be satisfied, optimization problem \eqref{OPn} is a structured stabilization problem subject to bounding constraints which is in general NP-hard \cite{toker1995np,blondel1997np}.

The selection of edge modification set $\mathcal{E}_s$ has a significant effect on the controllability-Gramian-based performance improvement achieved by the edge modification as edge modification matrix $\Delta(\gamma)$ appearing in the objective function \eqref{OPnobj} depends on edge modification set $\mathcal{E}_s$. For the low-cardinality edge modification scenario, i.e., a small value of $s$, we can assess how near-optimal the proposed ECM-based edge modification is. To that end, for a given $s \in \mathbb{N}_{|\mathcal{E}|}$, to identify the anti-optimal (i.e., the worst-case scenario ($\mathrm{WCS}$)) and optimal (i.e., the best-case scenario ($\mathrm{BCS}$)) edge modification sets $\mathcal{E}_s^{\mathrm{WCS}}$ and $\mathcal{E}_s^{\mathrm{BCS}}$ via a brute-force method, optimization problem \eqref{OPn} has to be solved for all $\binom{|\mathcal{E}|}{s}$ combinatorial choices of the edge modification set $\mathcal{E}_s$ which is computationally cumbersome (the NP-hardness). Then, for the large-scale network scenario, i.e., a large value of $N$, or the high-cardinality edge modification scenario, i.e., a large value of $s$, it is impossible to empirically assess the near-optimality performance of the ECM-based edge modification method compared to the worst-case and best-case scenarios' counterparts as its computational complexity is cumbersome. Including all the possible values for $s$, the computational complexity becomes $\mathcal{O}(2^{|\mathcal{E}|})$. Such a computational complexity issue motivates us to investigate the case in which $\mathcal{E}_s$ is chosen based on the first $s$ elements of the descending sorted sequence $\{\tau_l\}_{l=1}^{|\mathcal{E}|}$, i.e., $\{\tau_l\}_{l=1}^{s}$ with $\tau_1 \ge \dots \ge \tau_{s}$. Let us denote such $\mathcal{E}_s$ by $\mathcal{E}_s^{\mathrm{ECM}}$. We will solve optimization problem \eqref{OPn} associated with $\mathcal{E}_s^{\mathrm{ECM}}$ for $\gamma^{\mathrm{ECM}}$. Also, we will be able to measure the quality of the ECM-based modified network $A(L + \Delta (\gamma^{\mathrm{ECM}}))$ compared to a random modified network $A(L + \Delta (\gamma^{\mathrm{RND}}))$ obtained by solving optimization problem \eqref{OPn} associated with $\mathcal{E}_s^{\mathrm{RND}}$ for $\gamma^{\mathrm{RND}}$.

\subsection{Edge modification design via non-convex non-smooth optimization} \label{EMD2}

To convert the constrained optimization problem \eqref{OPn} to an unconstrained optimization problem while ensuring the satisfaction of $\alpha(A(L + \Delta(\gamma))) < 0$, $\|\gamma \| \le \beta$, and $\gamma \ge \iota_{\mathcal{E}_s}(L)$ in \eqref{StabCon}, \eqref{OPCn}, and \eqref{iotaL}, respectively, we can parameterize $\gamma \in \mathbb{R}^s$ as
\begin{align} \label{gChn}
    \gamma &= \beta \sin ({\pi \kappa}/{2}) {\nu}/{\|\nu\|},
\end{align}
with $\kappa \in \mathbb{R}$, $\nu \in \mathbb{R}^{s}$ and define $\mu(W_c(A(\mathcal{L})))$ as
\begin{align*}
    & \mu(W_c(A(\mathcal{L}))) := h(W_c(A(\mathcal{L}))) \times \texttt{r} - \xi \times (\neg \texttt{r}),\\ & \texttt{p} := \alpha(A(\mathcal{L})) < 0,~\texttt{q} := \gamma \ge \iota_{\mathcal{E}_s}(L),~\texttt{r} := \texttt{p} \land \texttt{q},
\end{align*}where $\xi$ is a large number acting as a practical infinity, e.g., $\xi = 10^{10}$. Then, utilizing $\eta := \begin{bmatrix}
    \nu^\top & \kappa 
\end{bmatrix}^\top \in \mathbb{R}^{s+1}$, we obtain the following equivalent unconstrained optimization problem:
\begin{align} \label{EOPn}
    \underset{\eta \in \mathbb{R}^{s+1}}{\max}&~\mu(W_c(A(L + \Delta(\eta)))).
\end{align}
To solve \eqref{EOPn} for $\eta$, one may utilize off-the-shelf nonlinear optimization solvers like MATLAB built-in function \texttt{fminsearch}. \texttt{fminsearch} solver has been developed based on Nelder--Mead simplex method \cite{lagarias1998convergence}. Given a non-convex non-smooth optimization problem, finding a globally optimal solution is generally a challenging task. We thus utilize the term \textit{near-optimal} instead of the term \textit{optimal}.

To parameterize $\gamma \in \mathbb{R}^s$ satisfying $\|\gamma\| \le \beta$ in \eqref{OPCn}, other than the sinusoidal term $\sin (\frac{\pi \kappa}{2})$ utilized in \eqref{gChn}, one can alternatively use any term with an upper-bound of $1$ on its absolute value, e.g., the sigmoid-type logistic term as $\frac{1}{1+e^{-\chi \kappa}}$ with an arbitrarily chosen $\chi > 0$. Note that selecting various parameterizations may lead to different levels of near-optimality for the solutions of \eqref{EOPn}.

\subsection{Edge modification design via convex optimization} \label{EMD3}

    Applying the Schur complement \cite{zhang2006schur} to $\gamma^\top \gamma \le \beta^2$ (which is equivalent to $\|\gamma\| \le \beta$ in \eqref{OPCn}), we get the equivalent linear matrix inequality (LMI) $\begin{bmatrix}
        \beta I_s & \gamma\\\gamma^\top & \beta
    \end{bmatrix} \succeq 0$ and subsequently optimization problem \eqref{OPn} can be recast as an optimization problem with a convex objective function and a set of linear matrix/vector inequalities along with a matrix equation with bi-linear terms as follows:
   \begin{subequations} \label{cvxcst}
       \begin{align}
    &\underset{\gamma \in \mathbb{R}^{s},~W \in \mathbb{R}^{(n-1) \times (n-1)}}{\max}~h(W),\\
&\mathrm{subject~to} \notag \\
& A(L+\Delta(\gamma))W+WA(L+\Delta(\gamma))^\top + B B^\top = 0,~ W \succeq 0, \label{cvxcstW}\\
& \begin{bmatrix}
    \beta I_s & \gamma\\ \gamma^\top & \beta
\end{bmatrix} \succeq 0,\\
& \gamma \ge \iota_{\mathcal{E}_s}(L). 
    \end{align}
   \end{subequations}Note that  
{\begin{align*}
    & A(L+\Delta(\gamma)) = A(L) + \underbrace{ \begin{bmatrix}
        0 & 0 & 0\\
        -M^{-1}\Delta(\gamma)U & 0 & 0\\
        0 & 0 & 0
    \end{bmatrix}}_{:=~\Xi(\Delta(\gamma))}, 
\end{align*}}holds according to the affine dependency of $A(\mathcal{L})$ on $\mathcal{L}$. Then, due to the appearing bi-linear terms $\Xi(\Delta(\gamma))W$ and $W\Xi(\Delta(\gamma))^\top$ in the Lyapunov equation in \eqref{cvxcstW}, the recast optimization problem \eqref{cvxcst} is non-convex in general. Therefore, convex optimization techniques cannot be applied directly to solve it for an optimal solution.

However, one can employ a path-following (homotopy)-based convex optimization alternative approach \cite{hassibi1999path} to approximately solve optimization problem \eqref{cvxcst} with bi-linear terms for $W$. Such a homotopy-based method can easily be implemented and is based on linearizing the bi-linear terms and iteratively solving a sequence of semi-definite programs (SDPs). Since this method is locally solving for solutions, similar to any other local method, the quality of its solutions potentially relies on the quality of the initialization.

Note that $\Delta(.)$ and $\Xi(.)$ are both linear operators implying that $\Xi(\Delta(.))$ is a linear operator. Then, to approximately solve optimization problem \eqref{cvxcst} for $W$ via a homotopy-based method, we construct the following convex optimization problem:\begin{subequations} \label{cvxcsthy}
       \begin{align}
    &\underset{\delta \gamma \in \mathbb{R}^{s},~\delta W \in \mathbb{R}^{(n-1) \times (n-1)}}{\max}~h(W_0+\delta W),\\
&\mathrm{subject~to} \notag \\
& (A(L) +\Xi(\Delta(\gamma_0)))(W_0 + \delta W) + \Xi(\Delta(\delta \gamma))W_0 + \notag \\
& (W_0 + \delta W)(A(L) +\Xi(\Delta(\gamma_0)))^\top + W_0 \Xi(\Delta(\delta \gamma))^\top \notag\\
& + B B^\top = 0,~W_0 + \delta W \succeq 0, \label{HM2}\\
& \begin{bmatrix}
    \beta I_s & \gamma_0 + \delta \gamma \\ (\gamma_0 + \delta \gamma)^\top & \beta
\end{bmatrix} \succeq 0,\\
& \gamma_0 + \delta \gamma \ge \iota_{\mathcal{E}_s}(L),\\
& \begin{bmatrix}
    c^{\mathrm{ub}} \|W_0\| I_{n-1} & \delta W\\
    \delta W^\top & c^{\mathrm{ub}} \|W_0\| I_{n-1}
\end{bmatrix} \succeq 0, \label{HM5}
    \end{align}
   \end{subequations}where $(W_0,\gamma_0)$ and $(\delta W, \delta \gamma)$ represent the initial guesses and the optimization variables (iterative improvements), respectively and $c^{\mathrm{ub}}$ is a constant parameter satisfying $0 < c^{\mathrm{ub}} < 1$. Note that the Lyapunov equation in \eqref{HM2} results from \eqref{cvxcstW} by eliminating the bi-linear terms $\Xi(\Delta(\delta \gamma)) \delta W$ and $\delta W \Xi(\Delta(\delta \gamma))^\top$. The intuitive reason behind such elimination is that the spectral norm of these bi-linear terms is sufficiently small due to the nature of the proposed homotopy-based method. It is noteworthy that the LMI in \eqref{HM5} is equivalent to $\|\delta W\| \le c^{\mathrm{ub}} \|W_0\|$ \cite{recht2010guaranteed} that bounds the spectral norm of $\delta W$ to ensure that the perturbation is small, and the linearized approximation is valid \cite{hassibi1999path}. To be more precise, we highlight that optimization problem \eqref{cvxcsthy} is iteratively solved with $(\delta W^{(i)}, \delta \gamma^{(i)})$ denoting the solution at the $i$-th iteration. Note that the initial guesses $(W_0^{(i)},\gamma_0^{(i)})$ are updated at each iteration via the following rules: for iteration $0$, we set $(W_0^{(0)},\gamma_0^{(0)}) := (W_0,\gamma_0)$ and for iterations $i \ge 1$, we set $(W_0^{(i)},\gamma_0^{(i)}) := (W_0^{(i-1)}+\delta W^{(i-1)},\gamma_0^{(i-1)}+\delta \gamma^{(i-1)})$. Moreover, we have $(W^{(i)},\gamma^{(i)}) = (W_0^{(i)} + \delta W^{(i)},\gamma_0^{(i)} + \delta \gamma^{(i)})$ and the stopping criterion can be considered as $|h(W^{(i)})-h(W^{(i-1)})| \le \epsilon_d |h(W^{(i-1)})|$ wherein $0 < \epsilon_d \ll 1$ denotes the desired tolerance parameter. Since there is no theoretical guarantee for the satisfaction of the stopping criterion, we also consider an upper bound for the number of iterations, namely $i_{\max}$.

\subsection{An ECM-based edge modification design procedure} \label{EMD4}

    For general networks in the network theory, the candidate edges of $\mathcal{E}$ are twofold: \textit{(i)} $(i,j) \in \mathcal{E} \cap \mathcal{E}^L$, and \textit{(ii)} $(i,j) \in \mathcal{E} \setminus \mathcal{E}^L$. For any $(i,j) \in \mathcal{E} \cap \mathcal{E}^L$, the edge can be varied (commonly known as edge modification) or eliminated (edge sparsification) as a special case. For any $(i,j) \in \mathcal{E} \setminus \mathcal{E}^L$, the edge can be added (edge addition). However, in the case of power networks, the edges can only be varied (modified). In other words, edge sparsification or addition is not feasible. The choice of candidate edge set $\mathcal{E}$ affects the controllability improvement of the proposed near-optimal edge modification. The ECM-based edge modification design procedure for power networks can be summarized as Procedure \ref{pro:two}.

\setlength{\floatsep}{5pt}{
\begin{algorithm}[!ht]
\caption{\textbf{ECM-based Edge Modification.}}\label{pro:two}
\DontPrintSemicolon
\textbf{Input}: $A(L)$, $B$, $\mathcal{E}$, $s$, $\beta$, $c^{\mathrm{ub}}$, $\epsilon_d$, $i_{\max}$.

Compute $W_c$ via solving the Lyapunov equation \eqref{RLEn}.

\For{$i = 1:N$}{
\For{$j = 1:N$}{Evaluate $F_{ji}$ via \eqref{WG2n}.\\Solve the Lyapunov equation \eqref{WG1n} for $X_{ji}$.\\Compute $\upsilon_{ji}$ via \eqref{ECMCn}.
}}

Construct ECM $\Upsilon$.

Compute the edge perturbation impact matrix $\mathcal{I}$ \eqref{FIM}.

Construct the descending sorted sequence of $\mathcal{I}_{ji} = |\upsilon_{ji}|$ for all $(i,j) \in \mathcal{E}$, namely $\{\tau_l\}_{l=1}^{|\mathcal{E}|}$ with $\tau_1 \ge \dots \ge \tau_{|\mathcal{E}|}$.

Construct the ECM-based edge modification set $\mathcal{E}_s^{\mathrm{ECM}}$ based on the first $s$ elements of the descending sorted sequence $\{\tau_l\}_{l=1}^{|\mathcal{E}|}$, i.e., $\{\tau_l\}_{l=1}^{s}$ with $\tau_1 \ge \dots \ge \tau_{s}$.

Choose the solver type:

\textbf{Solver type}: \textit{Non-convex non-smooth optimization solver}

Solve optimization problem \eqref{EOPn} for $\eta^{\mathrm{ECM}} \in \mathbb{R}^{s+1}$.

Compute $\gamma^{\mathrm{ECM}} \in \mathbb{R}^s$ via \eqref{gChn}.

\textbf{Solver type}: \textit{Convex optimization solver}

Solve optimization problem \eqref{cvxcsthy} for $\delta \gamma^{\mathrm{ECM}} \in \mathbb{R}^s$ and $\delta W^{\mathrm{ECM}} \in \mathbb{R}^{(n-1) \times (n-1)}$.

Compute $\gamma^{\mathrm{ECM}} \in \mathbb{R}^s$ via $\gamma^{\mathrm{ECM}} = \gamma^{\mathrm{ECM}}_0+ \delta \gamma^{\mathrm{ECM}}$.

Construct $\Delta(\gamma^{\mathrm{ECM}}) := \sum_{(i,j) \in \mathcal{E}_s^{\mathrm{ECM}}} \gamma_{k_{ji}}^{\mathrm{ECM}} V_{ji}$.

Compute $A(L+\Delta(\gamma^{\mathrm{ECM}}))$.

\textbf{Output}: $A(L+\Delta(\gamma^{\mathrm{ECM}}))$.

\end{algorithm}}

For a modified network $A(L+\Delta(\gamma))$, let us denote the corresponding modified (Kron-reduced) admittance matrix by $\hat{Y} = [\hat{y}_{ji}]$. Considering $l_{ji} - \gamma_{k_{ji}} = -|\hat{y}_{ji}|\mathbb{E}_j\mathbb{E}_i \cos (\theta_j^{(e)}-\theta_i^{(e)}-\hat{\varphi}_{ji})$, $\hat{\varphi}_{ji} = -\sign(j-i) \arctan \Big(\frac{\Re(\hat{y}_{ji})}{\Im(\hat{y}_{ji})}\Big)$ and defining a free (design) parameter $\varrho_{ji} := \tan (|\hat{\varphi}_{ji}|)$, one can obtain the following parameterized formula: 
{\begin{subequations} \label{ModAd}
\begin{align}
    \hat{y}_{ji} &= \Re (\hat{y}_{ji}) + \jmath \Im (\hat{y}_{ji}),\\
    \Re (\hat{y}_{ji}) &= \frac{\frac{\varrho_{ji}}{\sqrt{\varrho_{ji}^2+1}} (\gamma_{k_{ji}}-l_{ji})}{\mathbb{E}_j\mathbb{E}_i \cos (\theta_j^{(e)}-\theta_i^{(e)} + \sign (j-i) \arctan (\varrho_{ji}))},\\
    \Im (\hat{y}_{ji}) &= \frac{\frac{1}{\sqrt{\varrho_{ji}^2+1}} (\gamma_{k_{ji}}-l_{ji})}{\mathbb{E}_j\mathbb{E}_i \cos (\theta_j^{(e)}-\theta_i^{(e)} + \sign (j-i) \arctan (\varrho_{ji}))},
\end{align}    
\end{subequations}for the modified admittance matrix $\hat{Y}$. Interestingly, \eqref{ModAd} provides a systematic way to compute the modified admittance matrix $\hat{Y}$ in terms of the edge modification vector $\gamma$ obtained from Procedure \ref{pro:two}. In other words, by taking advantage of \eqref{ModAd}, we can effectively modify/redesign the power network. Moreover, \eqref{ModAd} facilitates formulating optimization problems with $\varrho_{ji}$s as optimization variables. A thorough investigation of such optimization problems is beyond the scope of this paper and can be considered a pertinent research direction.}

Furthermore, we highlight that changing the susceptance of transmission lines to perform the edge modifications, can be expensive because it often requires advanced technologies such as FACTS devices. These devices, allow precise control of transmission line properties but are costly to install and maintain. While it is important to consider the financial aspects of using such devices, this study focuses on the theoretical methods for improving power system controllability, not the benefit-cost analysis (BCA). Future research will explore practical applications, including conducting a detailed BCA, system reliability, and congestion reduction. This paper aims to build a strong theoretical foundation supported by detailed simulation studies.

\section{Case Studies and Research Questions} \label{secNS}

In this section, we verify the effectiveness of the ECM-based edge modification design procedure summarized by Procedure \ref{pro:two} via various IEEE power network benchmarks: IEEE $9$-bus benchmark ($N = 3$), IEEE $14$-bus benchmark ($N = 5$), and IEEE $68$-bus benchmark ($N = 16$). Throughout this section, the superscripts $\mathrm{WCS}$ and $\mathrm{BCS}$ refer to the worst-case and best-case scenarios, respectively. For comparative analysis, slightly modifying Procedure \ref{pro:two} (i.e., a different choice of $\mathcal{E}_s$), we compute $\gamma^{\mathrm{WCS}}$, $\gamma^{\mathrm{BCS}}$, and $\gamma^{\mathrm{RND}}$ via solving optimization problem \eqref{OPn} associated with $\mathcal{E}_s^{\mathrm{WCS}}$, $\mathcal{E}_s^{\mathrm{BCS}}$, and $\mathcal{E}_s^{\mathrm{RND}}$, respectively. To solve the continuous-time Lyapunov equation \eqref{CG} for $W_c$, we utilize the MATLAB built-in function $W_c = \texttt{lyap}(A(L),BB^\top)$. We empirically assess the controllability-Gramian-based performance improvement obtained by the ECM-based edge modification. For the low-cardinality edge modification scenario, we also evaluate how near-optimal the proposed ECM-based edge modification is. For simplicity, we only take into account the rotor angles $\theta(t) \in \mathbb{R}^N$ and rotor frequencies $\omega(t) \in \mathbb{R}^N$, and fast electrical dynamics $r(t) \in \mathbb{R}^{n-2N}$ are neglected. Then, $n = 2N$ holds for all the numerical experiments. In running Procedure \ref{pro:two} for the numerical experiments, the \textit{default} solver type we have chosen is the non-convex non-smooth optimization solver unless otherwise we explicitly mention the solver type is the convex optimization solver. All the numerical experiments have been conducted in MATLAB R$2024$a on a MacBook Pro with a $3.1$ GHz Intel Core i$5$ and memory $8$ GB $2133$ MHz.

To test the potential superiority of the dynamics-based centrality measures over the static alternatives, we additionally assess the relative performance of the proposed ECM-based method compared to a static graph-theoretic edge centrality-based alternative, namely nearest neighbor edge centrality (NNEC) \cite{brohl2022straightforward}, in identifying the network's most influential edges. The NNEC-based matrix $\Lambda = [\lambda_{ji}]$ is defined as follows \cite{brohl2022straightforward}:
{\begin{align*}
    \lambda_{ji} := \frac{\rho_j+\rho_i-2g_{ji}}{|\rho_j-\rho_i|+1}g_{ji}, 
\end{align*}}where $\rho_k$ denote the \textit{strength} of node $k$ defined as $\rho_k := \sum_{l} g_{kl}$. Intuitively, such a static graph-theoretic edge centrality identifies an edge to be more influential the larger its weight ($g_{ji}\uparrow$) and the more similar ($|\rho_j-\rho_i| \downarrow$) and the higher the strengths ($\rho_j+\rho_i\uparrow$) of the nodes that are connected by that edge \cite{brohl2022straightforward}. 

Throughout this section, we try to find appropriate answers to the following research questions:

{\begin{itemize}
    \item \textit{Q1}: Can we take advantage of the control-theoretic notion of edge centrality matrix (ECM) to strengthen the controllability of power networks via an edge modification procedure?
    \item \textit{Q2}: If the answer to \textit{Q1} is yes, compared to the best-case and worst-case scenarios, how near-optimal is the proposed ECM-based edge modification?
    \item \textit{Q3}: In the case of computationally prohibitive circumstances (e.g., the large-scale network or high-cardinality modification), can the proposed edge modification still be useful?
    \item \textit{Q4}: Do various Gramian-based performance metrics achieve the same controllability improvement?
    \item \textit{Q5}: How does the budget/energy upper-bound affect the controllability performance improvement?
    \item \textit{Q6}: How is the relative performance of the proposed ECM-based method compared to the static graph-theoretic edge centrality-based alternatives in identifying the network's most influential edges?
\end{itemize}}

\subsection{Edge modification computation example}

Let us consider the power system dynamics model \eqref{PNSSReduced} associated with the IEEE $9$-bus benchmark consisting of $N = 3$ generators with $M = \mathbf{diag}(\begin{bmatrix}
    0.1254 & 0.0340 & 0.0160
\end{bmatrix}^\top)$, $D = \mathbf{diag}(\begin{bmatrix}
    0.0125 & 0.0068 & 0.0048
\end{bmatrix}^\top)$, and $L$ as
{\begin{align*}
    L &= \begin{bmatrix}
    2.1276  & -0.9498  & -1.1778\\
   -0.9498  &  2.6715  & -1.7217\\
   -1.1778  & -1.7217  &  2.8995
    \end{bmatrix},
\end{align*}}via the Kron-reduction. Choosing $\mathcal{E} = \mathcal{E}^{L}$, $s = 2$, { $\beta = 1$ ($\beta > \min\{g_{ji}: (i,j) \in \mathcal{E} \}$),} and running Procedure \ref{pro:two}, we obtain $\Delta(\gamma^{\mathrm{ECM}_1})$, $\Delta(\gamma^{\mathrm{ECM}_2})$, and $\Delta(\gamma^{\mathrm{ECM}_3})$ as
{\begin{subequations} \label{ECMexam}
\begin{align}
    \Delta(\gamma^{\mathrm{ECM}_1}) &= \begin{bmatrix}
    -0.6134  &  0.9438  & -0.3304\\
    0.9438  & -0.9438    &     0\\
   -0.3304     &    0  &  0.3304
    \end{bmatrix},\\
    \Delta(\gamma^{\mathrm{ECM}_2}) &= \begin{bmatrix}
    -1.4141  &  0.7152  &  0.6989\\
    0.7152  & -0.7152    &     0\\
    0.6989    &     0  & -0.6989
    \end{bmatrix},\\
    \Delta(\gamma^{\mathrm{ECM}_3}) &= \begin{bmatrix}
    -1.4137  &  0.7258  &  0.6879\\
    0.7258  & -0.7258    &     0\\
    0.6879    &     0  & -0.6879
    \end{bmatrix}.
\end{align}
\end{subequations}}Note that imposing the condition $\beta > \min\{g_{ji}: (i,j) \in \mathcal{E}\}$ ensures that \eqref{iotaL} is not trivially (automatically) satisfied. In other terms, such an imposition prunes a trivial scenario. Equivalently, if we choose $\beta \le \min\{g_{ji}: (i,j) \in \mathcal{E}\}$, then \eqref{iotaL} becomes a redundant constraint. Observe that according to \eqref{OPCn} we have $|\gamma_{k_{ji}}| \le \|\gamma\| \le \beta$ for all $(i,j) \in \mathcal{E}_s$ and if $\beta \le \min\{g_{ji}: (i,j) \in \mathcal{E}\}$ holds, then we get $\beta \le g_{ji}$ for all $(i,j) \in \mathcal{E}$. Thus, considering $\mathcal{E}_s \subseteq \mathcal{E}$ and combining these inequalities, we get $-g_{ji} \le \gamma_{k_{ji}} \le g_{ji}$ for all $(i,j) \in \mathcal{E}_s$. Specifically, $\gamma_{k_{ji}} + g_{ji} \ge 0$ holds for all $(i,j) \in \mathcal{E}_s$, that is, \eqref{iotaL} is trivially (automatically) satisfied.
\subsection{Performance improvement measures}
To measure the performance improvement for any of the controllability-Gramian-based performance metrics in \eqref{PerfoMet}, we define the following performance measure $J(\gamma)~(\%)$:
{\begin{align} \label{PIn}
& J(\gamma) = 100 \times \frac{h(W_c(A(L+\Delta(\gamma))))-h(W_c(A(L)))}{|h(W_c(A(L)))|},
\end{align}}Furthermore, let us define the near-optimality performance measures $J_V(\gamma^{\mathrm{ECM}})~(\%)$ (value-based) and $J_C(\gamma^{\mathrm{ECM}})~(\%)$ (cardinality-based) as
{\begin{subequations} \label{JVJC}
\begin{align}
  &J_V(\gamma^{\mathrm{ECM}}) := 100 \times \frac{J(\gamma^{\mathrm{ECM}})-J(\gamma^{\mathrm{WCS}})}{J(\gamma^{\mathrm{BCS}})-J(\gamma^{\mathrm{WCS}})}, \label{JVl}\\
    &J_C(\gamma^{\mathrm{ECM}}) := 100 \times \frac{\Big | \Big \{i \in \mathbb{N}_{\binom{|\mathcal{E}|}{s}}: J_i \le J(\gamma^{\mathrm{ECM}}) \Big\}\Big|}{\binom{|\mathcal{E}|}{s}},\label{JCl}
\end{align}    
\end{subequations}}where $J_i$ represents the performance improvement associated with the $i$-th combination out of the $\binom{|\mathcal{E}|}{s}$ total scenarios. Also, observe that the following inequalities hold by definition: $J(\gamma^{\mathrm{WCS}}) \le J(\gamma^{\mathrm{ECM}}) \le J(\gamma^{\mathrm{BCS}})$. For $i \in \{1,2,3\}$, similar to $\Delta(\gamma^{\mathrm{ECM}_i})$ in \eqref{ECMexam}, we can compute $\Delta(\gamma^{\mathrm{WCS}_i})$ and $\Delta(\gamma^{\mathrm{BCS}_i})$ via running the slightly modified Procedure \ref{pro:two} for the corresponding $\mathcal{E}_s^{\mathrm{WCS}_i}$ and $\mathcal{E}_s^{\mathrm{BCS}_i}$, respectively.

Considering all the IEEE benchmarks and their corresponding $M$, $D$, and $\beta$, we re-run Procedure \ref{pro:two} and its slightly modified version for $s = 1$. Tab. \ref{tab:my_label1} reflects the performance improvement value $J(\gamma)~(\%)$ defined by \eqref{PIn} associated with the ECM-based, WCS, and BCS for various choices of the controllability-Gramian-based performance metrics in \eqref{PerfoMet} and $s \in \{1,2\}$. As shown in Tab. \ref{tab:my_label1}, as expected, we observe that the fewer edge modifications we have, the less performance improvement we get. Built upon the data reflected by Tab. \ref{tab:my_label1}, we obtain Tab. \ref{tab:my_label2} that demonstrates the near-optimality values $J_V(\gamma^{\mathrm{ECM}})$ and $J_C(\gamma^{\mathrm{ECM}})$ associated with $\Delta(\gamma^{\mathrm{ECM}})$ for various choices of the Gramian-based performance metrics in \eqref{PerfoMet} and $s \in \{1,2\}$. Moreover, Tab. \ref{tab:my_label3} specifies the corresponding $\mathcal{E}_s^{\mathrm{WCS}}$, $\mathcal{E}_s^{\mathrm{ECM}}$, and $\mathcal{E}_s^{\mathrm{BCS}}$, respectively. As illustrated by Tab. \ref{tab:my_label1}, except for the first Gramian-based performance metric, i.e., the trace performance metric, the other Gramian-based performance metrics achieve acceptable near-optimality performances.

\begin{table}[t]
    \centering
        \caption{Performance improvement value $J(\gamma)~(\%)$ defined by \eqref{PIn} associated with the WCS, the ECM-based scenario, and the BCS for various choices of the Gramian-based performance metrics in \eqref{PerfoMet} and $s \in \{1,2\}$.}
    {\begin{tabular}{|c|c|c|c|c|}
    \hline
        $s$ & \diagbox{$g(W_c)$}{$J(\gamma)$} & $J(\gamma^{\mathrm{WCS}})$ & $J(\gamma^{\mathrm{ECM}})$ & $J(\gamma^{\mathrm{BCS}})$ \\
        \hline
        \multicolumn{5}{|c|}{IEEE $9$-bus benchmark $(N = 3)$} \\
        \hline
        $1$ & $\mathrm{tr}(W_c)$ & $0.6012$ &   $0.6012$  &  $0.9853$ \\
         \hline
        $1$ & $\log \det(W_c)$ & $1.7967$ &   $3.1898$  &  $3.1898$ \\
         \hline
        $1$ & $-\mathrm{tr}(W_c^{-1})$ & $21.4248$  & $28.1474$  & $28.1474$ \\
        \hline
        $2$ & $\mathrm{tr}(W_c)$ & $0.7644$    & $0.7644$ &  $1.0913$ \\
         \hline
        $2$ & $\log \det(W_c)$ & $3.5371$  &  $4.5303$  &  $4.5303$ \\
         \hline
        $2$ & $-\mathrm{tr}(W_c^{-1})$ & $36.4843$  & $39.2109$  & $39.2109$ \\
        \hline
        \multicolumn{5}{|c|}{IEEE $14$-bus benchmark $(N = 5)$} \\
        \hline
        $1$ & $\mathrm{tr}(W_c)$ & $0.0137$ &   $0.4415$  &  $0.7511$ \\
         \hline
        $1$ & $\log \det(W_c)$ & $0.8121$ &   $3.8422$  &  $3.8422$ \\
         \hline
        $1$ & $-\mathrm{tr}(W_c^{-1})$ & $4.3888$  &  $8.9578$  &  $8.9578$ \\
        \hline
        $2$ & $\mathrm{tr}(W_c)$ & $0.1203$ &   $0.8782$  &  $0.9959$ \\
         \hline
        $2$ & $\log \det(W_c)$ & $1.5138$ &    $5.3361$  &  $6.0016$ \\
         \hline
        $2$ & $-\mathrm{tr}(W_c^{-1})$ & $6.9083$ &  $11.1605$ &  $11.8227$ \\
        \hline
        \multicolumn{5}{|c|}{IEEE $68$-bus benchmark $(N = 16)$} \\
        \hline
        $1$ & $\mathrm{tr}(W_c)$ & $0.0000$ & $0.0014$ & $0.0014$ \\
         \hline
        $1$ & $\log \det(W_c)$ & $0.0002$ & $0.0062$ & $0.0062$ \\
         \hline
        $1$ & $-\mathrm{tr}(W_c^{-1})$ & $0.0000$ & $0.0081$ & $0.0081$ \\
        \hline
        $2$ & $\mathrm{tr}(W_c)$ & $0.0000$ & $0.0019$ & $0.0019$ \\
         \hline
        $2$ & $\log \det(W_c)$ & $0.0005$ & $0.0084$ & $0.0084$ \\
         \hline
        $2$ & $-\mathrm{tr}(W_c^{-1})$ & $0.0000$ & $0.0113$ & $0.0113$ \\
        \hline
    \end{tabular}}
    \label{tab:my_label1}
\end{table}

\begin{table}[t]
    \centering
        \caption{Near-optimality values $J_V(\gamma^{\mathrm{ECM}})~(\%)$ and $J_C(\gamma^{\mathrm{ECM}})~(\%)$ associated with $\Delta(\gamma^{\mathrm{ECM}})$ for various choices of the Gramian-based performance metrics in \eqref{PerfoMet} and $s \in \{1,2\}$.}
    {\begin{tabular}{|c|c|c|c|}
    \hline
        $s$ & $g(W_c)$ & $J_V(\gamma^{\mathrm{ECM}})$ & $J_C(\gamma^{\mathrm{ECM}})$ \\
        \hline
        \multicolumn{4}{|c|}{IEEE $9$-bus benchmark $(N = 3)$} \\
        \hline
        $1$ & $\mathrm{tr}(W_c)$ & $0.00$ & $33.33$ \\
         \hline
        $1$ & $\log \det(W_c)$ & $100.00$ & $100.00$ \\
         \hline
        $1$ & $-\mathrm{tr}(W_c^{-1})$ & $100.00$ & $100.00$ \\
        \hline
        $2$ & $\mathrm{tr}(W_c)$  & $0.00$ & $33.33$  \\
         \hline
        $2$ & $\log \det(W_c)$ & $100.00$ & $100.00$ \\
         \hline
        $2$ & $-\mathrm{tr}(W_c^{-1})$ & $100.00$ & $100.00$ \\
        \hline
        \multicolumn{4}{|c|}{IEEE $14$-bus benchmark $(N = 5)$} \\
        \hline
        $1$ & $\mathrm{tr}(W_c)$ & $58.01$ & $71.43$ \\
         \hline
        $1$ & $\log \det(W_c)$ & $100.00$ & $100.00$ \\
         \hline
        $1$ & $-\mathrm{tr}(W_c^{-1})$ & $100.00$ & $100.00$ \\
        \hline
        $2$ & $\mathrm{tr}(W_c)$  & $86.55$ & $95.24$  \\
         \hline
        $2$ & $\log \det(W_c)$ & $85.17$ & $95.24$ \\
         \hline
        $2$ & $-\mathrm{tr}(W_c^{-1})$ & $86.52$ & $90.48$ \\
        \hline
        \multicolumn{4}{|c|}{IEEE $68$-bus benchmark $(N = 16)$} \\
        \hline
        $1$ & $\mathrm{tr}(W_c)$ & $100.00$ & $100.00$ \\
         \hline
        $1$ & $\log \det(W_c)$ & $100.00$ & $100.00$ \\
         \hline
        $1$ & $-\mathrm{tr}(W_c^{-1})$ & $100.00$ & $100.00$ \\
        \hline
        $2$ & $\mathrm{tr}(W_c)$  & $100.00$ & $100.00$ \\
         \hline
        $2$ & $\log \det(W_c)$ & $100.00$ & $100.00$ \\
         \hline
        $2$ & $-\mathrm{tr}(W_c^{-1})$ & $100.00$ & $100.00$ \\
        \hline
    \end{tabular}}
    \label{tab:my_label2}
\end{table}

\begin{table}[t]
    \centering
        \caption{Edge modification sets associated with the WCS, the ECM-based scenario, and the BCS for various choices of the Gramian-based performance metrics in \eqref{PerfoMet} and $s \in \{1,2\}$.}
    {\begin{tabular}{|c|c|c|c|}
    \hline
        $(s,i)$ & $\in \mathcal{E}_s^{\mathrm{WCS}_i}$ & $\in \mathcal{E}_s^{\mathrm{ECM}_i}$ & $\in \mathcal{E}_s^{\mathrm{BCS}_i}$ \\
        \hline
        \multicolumn{4}{|c|}{IEEE $9$-bus benchmark $(N = 3)$} \\
        \hline
        $(1,1)$ & $(3,1)$ & $(3,1)$ & $(3,2)$ \\
         \hline
        $(1,2)$ & $(3,2)$ & $(3,1)$ & $(3,1)$ \\
         \hline
        $(1,3)$ & $(3,2)$ & $(3,1)$ & $(3,1)$ \\
        \hline
        $(2,1)$ & $(2,1),(3,1)$ & $(2,1),(3,1)$ & $(2,1),(3,2)$ \\
        \hline
        $(2,2)$ & $(2,1),(3,2)$ & $(2,1),(3,1)$ & $(2,1),(3,1)$ \\
         \hline
        $(2,3)$ & $(3,1),(3,2)$ & $(2,1),(3,1)$ & $(2,1),(3,1)$ \\
        \hline
        \multicolumn{4}{|c|}{IEEE $14$-bus benchmark $(N = 5)$} \\
        \hline
        $(1,1)$ & $(5,4)$ & $(3,2)$ & $(5,1)$ \\
         \hline
        $(1,2)$ & $(5,4)$ & $(4,3)$ & $(4,3)$ \\
         \hline
        $(1,3)$ & $(5,1)$ & $(4,3)$ & $(4,3)$ \\
        \hline
        $(2,1)$ & $(5,2),(5,4)$ & $(3,2),(5,1)$ & $(4,2),(5,1)$ \\
         \hline
        $(2,2)$ & $(3,2),(5,4)$ & $(4,3),(5,1)$ & $(3,2),(4,3)$ \\
         \hline
        $(2,3)$ & $(5,1),(5,2)$ & $(4,2),(4,3)$ & $(3,2),(4,3)$ \\
        \hline
        \multicolumn{4}{|c|}{IEEE $68$-bus benchmark $(N = 16)$} \\
        \hline
        $(1,1)$ & $(12,3)$ & $(16,5)$ & $(16,5)$ \\
         \hline
        $(1,2)$ & $(13,11)$ & $(16,5)$ & $(16,5)$ \\
         \hline
        $(1,3)$ & $(12,3)$ & $(15,14)$ & $(15,14)$ \\
        \hline
        $(2,1)$ & $(3,2),(12,3)$ & $(11,5),(16,5)$ & $(11,5),(16,5)$ \\
         \hline
        $(2,2)$ & $(12,11),(13,11)$ & $(16,4),(16,5)$ & $(16,4),(16,5)$ \\
         \hline
        $(2,3)$ & $(11,7),(12,3)$ & $(15,14),(16,13)$ & $(15,14),(16,13)$ \\
        \hline
    \end{tabular}}
    \label{tab:my_label3}
\end{table}

For the low-cardinality edge modification scenario, we run Procedure \ref{pro:two} with the convex (C) solver to compare the dynamics-based solution and static alternative as reflected on Tab. \ref{tab:my_label0}. As depicted by Tab. \ref{tab:my_label0}, the proposed ECM-based method outperforms the NNEC-based method for almost all the cases. Thus, the dynamics-based centrality measure is more successful than the static alternative in identifying the network's most influential $s$ edges for the low-cardinality edge modification scenario.

\begin{table}[t]
    \centering
        \caption{Performance improvement value $J(\gamma)~(\%)$ defined by \eqref{PIn} associated with the WCS, the ECM-based scenario, the NNEC-based scenario, and the BCS for various choices of the Gramian-based performance metrics in \eqref{PerfoMet}, $s \in \{1,2\}$, and convex (C) solver.}
    {\begin{tabular}{|c|c|c|c|c|}
    \hline
        $(s,i)$ & $J(\gamma^{\mathrm{WCS}_i}_{\mathrm{C}})$ & $J(\gamma^{\mathrm{ECM}_i}_{\mathrm{C}})$ & $J(\gamma^{\mathrm{NNEC}_i}_{\mathrm{C}})$ & $J(\gamma^{\mathrm{BCS}_i}_{\mathrm{C}})$ \\
        \hline
        \multicolumn{5}{|c|}{IEEE $9$-bus benchmark $(N = 3)$} \\
        \hline
        $(1,1)$ & $0.6012$ & $0.6012$ & $\mathbf{0.9853}$ & $0.9853$\\
         \hline
        $(1,2)$ & $1.7967$ & $\mathbf{3.1898}$ & $1.7967$ & $3.1898$ \\
         \hline
        $(1,3)$ & $21.4248$ & $\mathbf{28.1474}$ & $21.4248$ & $28.1474$ \\
        \hline
        $(2,1)$ & $0.7644$ & $0.7644$ & $\mathbf{1.0912}$ & $1.0912$ \\
         \hline
        $(2,2)$ & $3.5371$ & $\mathbf{4.5303}$ & $3.5371$ & $4.5303$ \\
         \hline
        $(2,3)$ & $36.9827$ & $\mathbf{39.2109}$ & $36.9827$ & $39.2109$ \\
        \hline
        \multicolumn{5}{|c|}{IEEE $14$-bus benchmark $(N = 5)$} \\
        \hline
        $(1,1)$ & $0.0137$ & $\mathbf{0.4415}$ & $0.0137$ & $0.7511$ \\
         \hline
        $(1,2)$ & $0.8121$ & $\mathbf{3.8422}$ & $0.8121$ & $3.8422$ \\
         \hline
        $(1,3)$ & $4.3888$ & $\mathbf{8.9578}$ & $5.6821$ & $8.9578$ \\
        \hline
        $(2,1)$ & $0.0357$ & $\mathbf{0.8782}$ & $0.0357$ & $0.9985$ \\
         \hline
        $(2,2)$ & $1.5137$ & $\mathbf{5.3361}$ & $2.1620$ & $5.6098$ \\
         \hline
        $(2,3)$ &  $6.9083$ & $\mathbf{11.3180}$ & $7.9061$ & $11.8227$ \\
        \hline
        \multicolumn{5}{|c|}{IEEE $68$-bus benchmark $(N = 16)$} \\
        \hline
        $(1,1)$ & $0.0000$ & $\mathbf{0.0014}$ & $0.0000$ & $0.0014$ \\
         \hline
        $(1,2)$ & $0.0002$ & $\mathbf{0.0062}$ & $0.0004$ & $0.0062$ \\
         \hline
        $(1,3)$ & $0.0000$ & $\mathbf{0.0081}$ & $0.0012$ & $0.0081$ \\
        \hline
        $(2,1)$ & $0.0000$ & $\mathbf{0.0019}$ & $0.0000$ & $0.0019$ \\
         \hline
        $(2,2)$ & $0.0005$ & $\mathbf{0.0084}$ & $0.0012$ & $0.0084$ \\
         \hline
        $(2,3)$ & $0.0001$ & $\mathbf{0.0113}$ & $0.0013$ & $0.0113$ \\
        \hline
    \end{tabular}}
    \label{tab:my_label0}
\end{table}

\subsection{Near-optimal topologies' visualization}

\begin{figure*}[!ht]
    \centering
    \includegraphics[scale = 0.32,trim={3cm 1.5cm 3cm 0},clip]{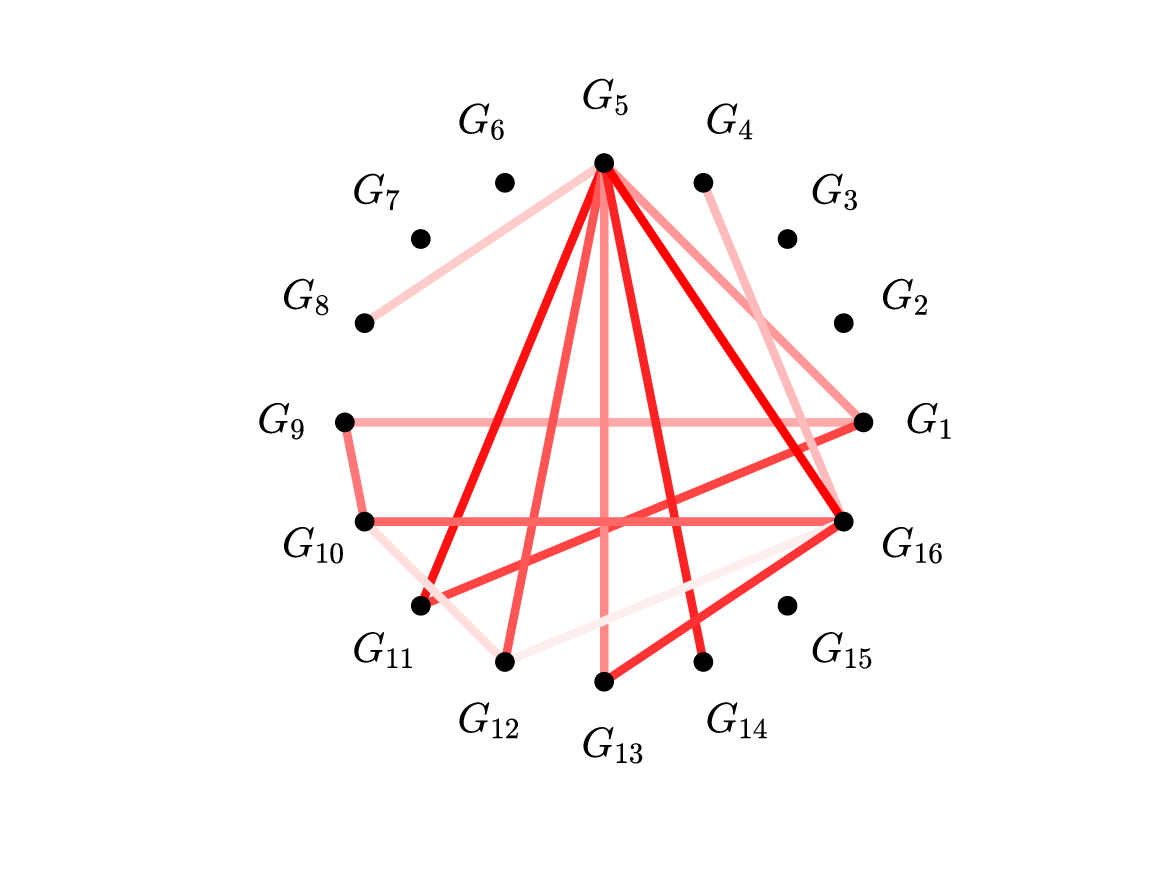}
    \includegraphics[scale = 0.32,trim={3cm 1.5cm 3cm 0},clip]{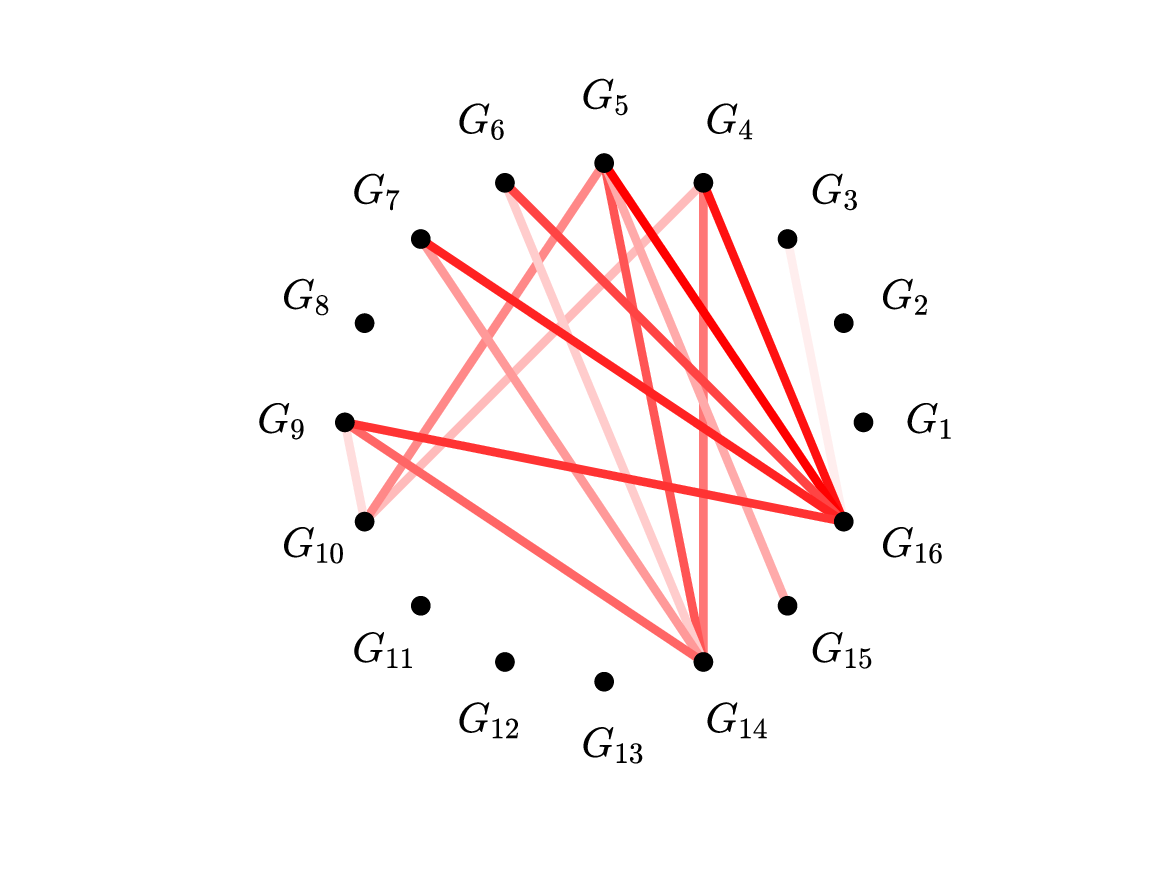}
    \includegraphics[scale = 0.32,trim={3cm 1.5cm 3cm 0},clip]{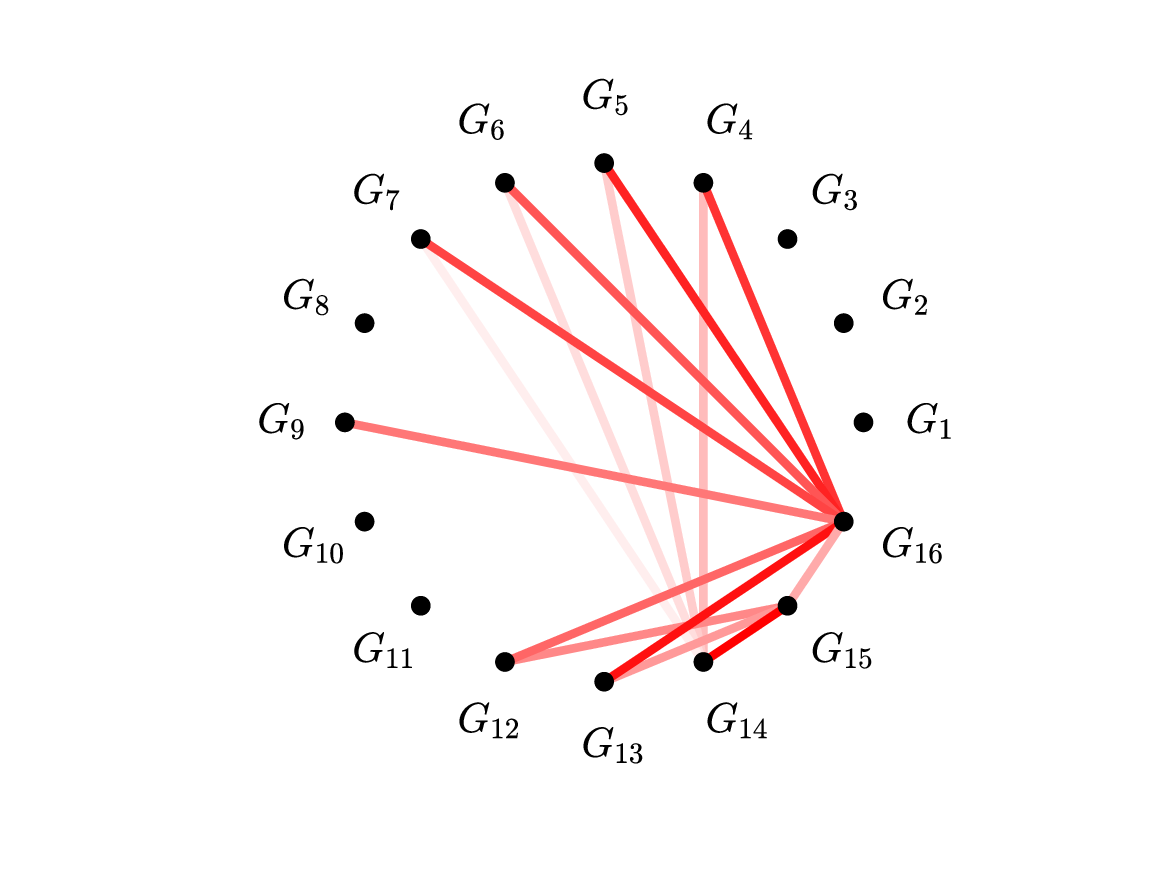}
    \includegraphics[scale = 0.32,trim={3cm 1.5cm 3cm 0},clip]{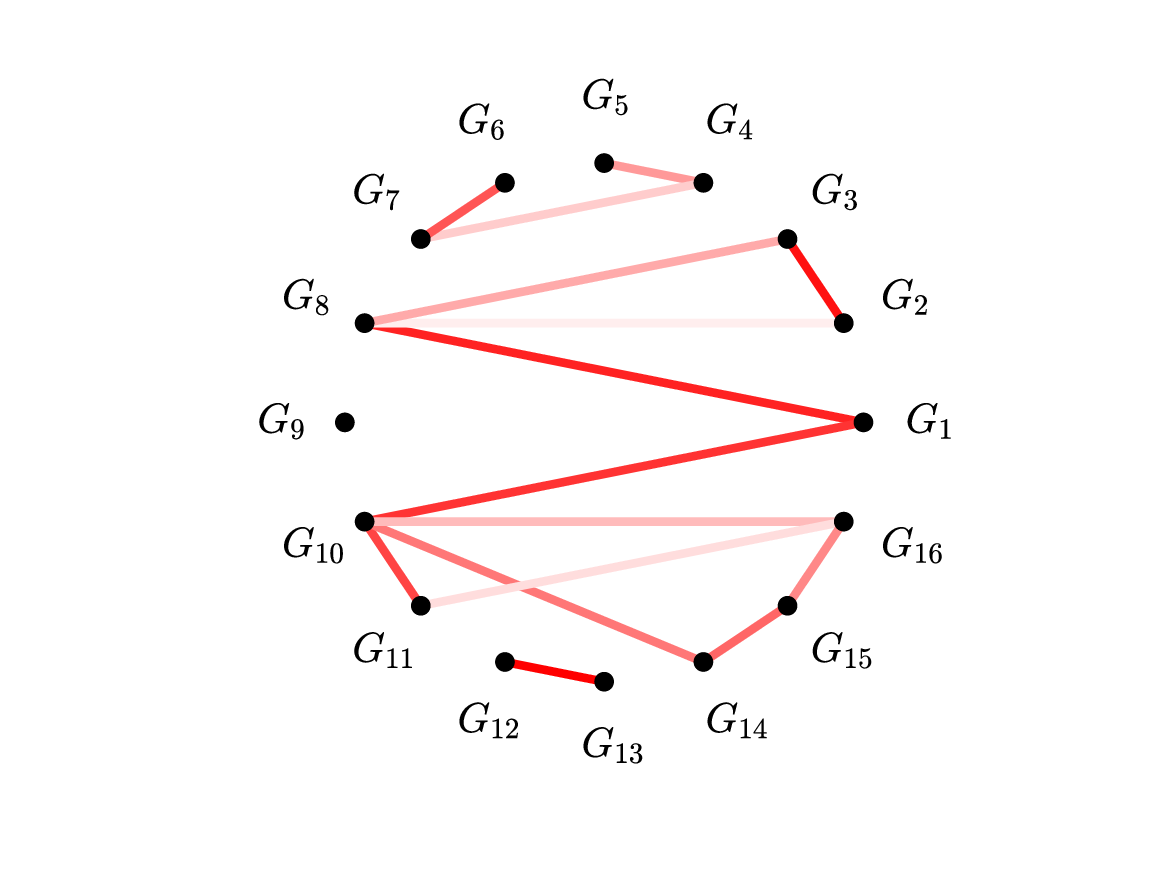}
\caption{Near-optimal topologies of $\Delta(\gamma^{\mathrm{ECM}_1})$, $\Delta(\gamma^{\mathrm{ECM}_2})$, $\Delta(\gamma^{\mathrm{ECM}_3})$ and the topology of $\Delta(\gamma^{\mathrm{NNEC}})$ (from left to right, respectively) where spectrum red lines represent the modified edges. The edges with darker colors correspond to the more influential edges. IEEE $68$-bus benchmark with $N = 16$ generators and $s=15$.}
    \label{fig:enter-label}
\end{figure*}

Choosing $\mathcal{E} = \mathcal{E}^{L}$, $s = 15$, { $\beta = 0.0024$ ($\beta > \min\{g_{ji}: (i,j) \in \mathcal{E}\}$),} and running Procedure \ref{pro:two}, we visualize the near-optimal topologies of $\Delta(\gamma^{\mathrm{ECM}_1})$ (first), $\Delta(\gamma^{\mathrm{ECM}_2})$ (second), and $\Delta(\gamma^{\mathrm{ECM}_3})$ (third) in Fig. \ref{fig:enter-label} for IEEE $68$-bus benchmark with $N = 16$ generators and $s=15$. Since $\binom{120}{15} > 4.7305 \times 10^{18}$ holds, the computation of $J(\gamma^{\mathrm{WCS}})$, $J(\gamma^{\mathrm{BCS}})$, $J_V(\gamma^{\mathrm{ECM}})$, and $J_C(\gamma^{\mathrm{ECM}})$ is impossible while the edge centrality notion facilitates computing the $\gamma^{\mathrm{ECM}_i}$ for $i \in \{1,2,3\}$ with $J(\gamma^{\mathrm{ECM}_1}) = 0.0030~\%$, $J(\gamma^{\mathrm{ECM}_2}) = 0.0160~\%$, and $J(\gamma^{\mathrm{ECM}_3}) = 0.0220~\%$. Running the slightly modified version of Procedure \ref{pro:two} for a randomly constructed $\mathcal{E}_s^{\mathrm{RND}}$ with $s = 15$, we obtain the $\gamma^{\mathrm{RND}}_i$ for $i \in \{1,2,3\}$ with $J(\gamma^{\mathrm{RND}_1}) = 0.0018~\%$, $J(\gamma^{\mathrm{RND}_2}) = 0.0084~\%$, and $J(\gamma^{\mathrm{RND}_3}) = 0.0081~\%$. A simple comparison $\begin{bmatrix}
    0.0030 & 0.0160 & 0.0220
\end{bmatrix}^\top > \begin{bmatrix}
    0.0018 & 0.0084 & 0.0081
\end{bmatrix}^\top$ shows that the ECM-based edge modification outperforms the random-based alternative in terms of performance improvement by $66.67~\%$, $90.48~\%$, and $171.60~\%$ relative superiority, respectively. Unsurprisingly, such superiority is mainly due to the performance sensitivity information captured by the edge centrality notion. Again, we observe that the first Gramian-based performance metric has the weakest functionality compared to the other Gramian-based performance metrics.

Repeating the numerical experiment with the convex optimization solver, we obtain the results reflected on Tab. \ref{tab:my_label22}. As shown by Tab. \ref{tab:my_label22}, compared to the non-convex solver, the convex solver obtains better performance improvements (about $10~\%$ in some items) at the expense of more computational time \cite{majumdar2020recent}. For the low-cardinality edge modification scenario in Tab. \ref{tab:my_label1}, a similar comprehensive re-examination reveals that the non-convex solver is faster than the convex alternative while attaining almost identical performance improvements.

\begin{table}[t]
    \centering
        \caption{Performance improvement value $J(\gamma)~(\%)$ defined by \eqref{PIn} associated with the ECM-based scenario and the Random scenario for various choices of the Gramian-based performance metrics in \eqref{PerfoMet}, IEEE $68$-bus benchmark, $s = 15$, and both solvers (non-convex (NC) and convex (C)).}
    {\begin{tabular}{|c|c|c|c|c|}
    \hline
        \diagbox{$g(W_c)$}{$J(\gamma)$} & $J(\gamma^{\mathrm{ECM}}_{\mathrm{NC}})$ & $J(\gamma^{\mathrm{RND}}_{\mathrm{NC}})$ & $J(\gamma^{\mathrm{ECM}}_{\mathrm{C}})$ & $J(\gamma^{\mathrm{RND}}_{\mathrm{C}})$\\
        \hline
        $\mathrm{tr}(W_c)$ & $0.0030$ & $0.0018$ & $\mathbf{0.0033}$ & $\mathbf{0.0019}$ \\
        \hline
        $\log \det(W_c)$ & $0.0160$ & $0.0084$ & $\mathbf{0.0167}$ & $\mathbf{0.0095}$ \\
        \hline
        $-\mathrm{tr}(W_c^{-1})$ & $0.0220$ & $0.0081$ & $\mathbf{0.0238}$ & $\mathbf{0.0083}$ \\
        \hline
    \end{tabular}}

    \label{tab:my_label22}
\end{table}

Tab. \ref{tab:my_label3} shows the edge modification sets associated with the WCS, the ECM-based scenario, and the BCS for various choices of the Gramian-based performance metrics in \eqref{PerfoMet} and $s \in \{1,2\}$. Also, the edge modification sets associated with the NNEC-based scenario for various IEEE benchmarks and $s \in \{1,2\}$ are as follows: \textit{(i)} IEEE $9$-bus benchmark: $\mathcal{E}_s^{\mathrm{NNEC}} = \{(3,2)\}$ for $s=1$ and $\mathcal{E}_s^{\mathrm{NNEC}} = \{(2,1),(3,2)\}$ for $s=2$, \textit{(ii)} IEEE $14$-bus benchmark: $\mathcal{E}_s^{\mathrm{NNEC}} = \{(5,4)\}$ for $s=1$ and $\mathcal{E}_s^{\mathrm{NNEC}} = \{(5,2),(5,4)\}$ for $s=2$, and \textit{(iii)} IEEE $68$-bus benchmark: $\mathcal{E}_s^{\mathrm{NNEC}} = \{(13,12)\}$ for $s=1$ and $\mathcal{E}_s^{\mathrm{NNEC}} = \{(3,2),(13,12)\}$ for $s=2$. As a limitation of the NNEC-based alternative, note that the NNEC measure is unable to distinguish various choices of the Gramian-based performance metrics from each other. In other terms, it proposes the same edge modification set for all Gramian-based performance metrics. As depicted in Tab. \ref{tab:my_label3} and the edge modification sets associated with the NNEC-based scenario, the ECM approach has mostly identified the optimal topologies, leading to a better near-optimality level. Similar to the earlier observations, the first Gramian-based performance metric has the weakest functionality compared to the other Gramian-based performance metrics.

Likewise, for the high-cardinality edge modification scenario, we run Procedure \ref{pro:two} with the convex (C) solver to compare the dynamics-based solution and static alternative as reflected on Tab. \ref{tab:my_labeln1}. As depicted by Tab. \ref{tab:my_labeln1}, similarly, the proposed ECM-based method outperforms the NNEC-based method for all the cases. Thus, the dynamics-based centrality measure is similarly more successful than the static alternative in identifying the network's most influential $s$ edges for the high-cardinality edge modification scenario. Also, we visualize the topology of $\Delta(\gamma^{\mathrm{NNEC}})$ (fourth) in Fig. \ref{fig:enter-label}.

\begin{table}[t]
    \centering
        \caption{Performance improvement value $J(\gamma)~(\%)$ defined by \eqref{PIn} associated with the ECM-based scenario, the NNEC-based scenario, and the Random scenario for various choices of the Gramian-based performance metrics in \eqref{PerfoMet}, IEEE $68$-bus benchmark, $s = 15$, and convex (C) solver.}
    {\begin{tabular}{|c|c|c|c|}
    \hline
        \diagbox{$g(W_c)$}{$J(\gamma)$} &
        $J(\gamma^{\mathrm{ECM}}_{\mathrm{C}})$ &
        $J(\gamma^{\mathrm{NNEC}}_{\mathrm{C}})$ &
        $J(\gamma^{\mathrm{RND}}_{\mathrm{C}})$\\
        \hline
        $\mathrm{tr}(W_c)$ & $\mathbf{0.0033}$ & $0.0013$ & $0.0019$ \\
        \hline
        $\log \det(W_c)$ & $\mathbf{0.0167}$ & $0.0058$ & $0.0095$ \\
        \hline
        $-\mathrm{tr}(W_c^{-1})$ & $\mathbf{0.0238}$ & $0.0112$ & $0.0083$ \\
        \hline
    \end{tabular}}

    \label{tab:my_labeln1}
\end{table}

\subsection{Minimum energy control quality assessment}

For an IEEE benchmark, the minimum energy required to steer the state from $x_0$ to $x_f = 0$ in $t_f$ time unit is equal to $x_0^\top W_c(t_f)^{-1} x_0$. The minimum energy control input has the following form \cite{chen1984linear}:
{\begin{align} \label{umin}
    u(t) &= -B^\top e^{A^\top(t_f-t)} W_c(t_f)^{-1} e^{At_f}x_0,
\end{align}}for which the corresponding minimum energy is equal to $J_u := x_0^\top W_c(t_f)^{-1} x_0$. It can be verified that for $x_0 \in \mathcal{N}(0,I)$, the average (mean) minimum energy associated with the minimum energy control input in \eqref{umin} becomes 
{\begin{align*}
    & \mathbf{E}(x_0^\top W_c(t_f)^{-1} x_0) = \mathbf{E}(\mathrm{tr}(x_0 x_0^\top W_c(t_f)^{-1})) =\\
    & \mathrm{tr}(\mathbf{E}(x_0 x_0^\top) W_c(t_f)^{-1}) = \mathrm{tr}(W_c(t_f)^{-1}),
\end{align*}}as $\mathrm{tr}(x_0^\top W_c(t_f)^{-1}x_0) = \mathrm{tr}(x_0 x_0^\top W_c(t_f)^{-1})$ holds, the expectation and trace operators commute, and for $x_0 \in \mathcal{N}(0,I)$, $\mathbf{E}(x_0 x_0^\top)$ holds. To compute $W_c(t_f)$ in \eqref{CG}, one can solve the following continuous-time Lyapunov equation:
{\begin{align} \label{ALCG}
    & A(\mathcal{L}) W_c(t_f) + W_c(t_f) A(\mathcal{L})^\top + B B^\top \notag\\
    & - e^{At_f}B B^\top e^{A^\top t_f} = 0.
\end{align}}

\begin{figure*}[t]
    \centering
    \includegraphics[scale = 0.3]{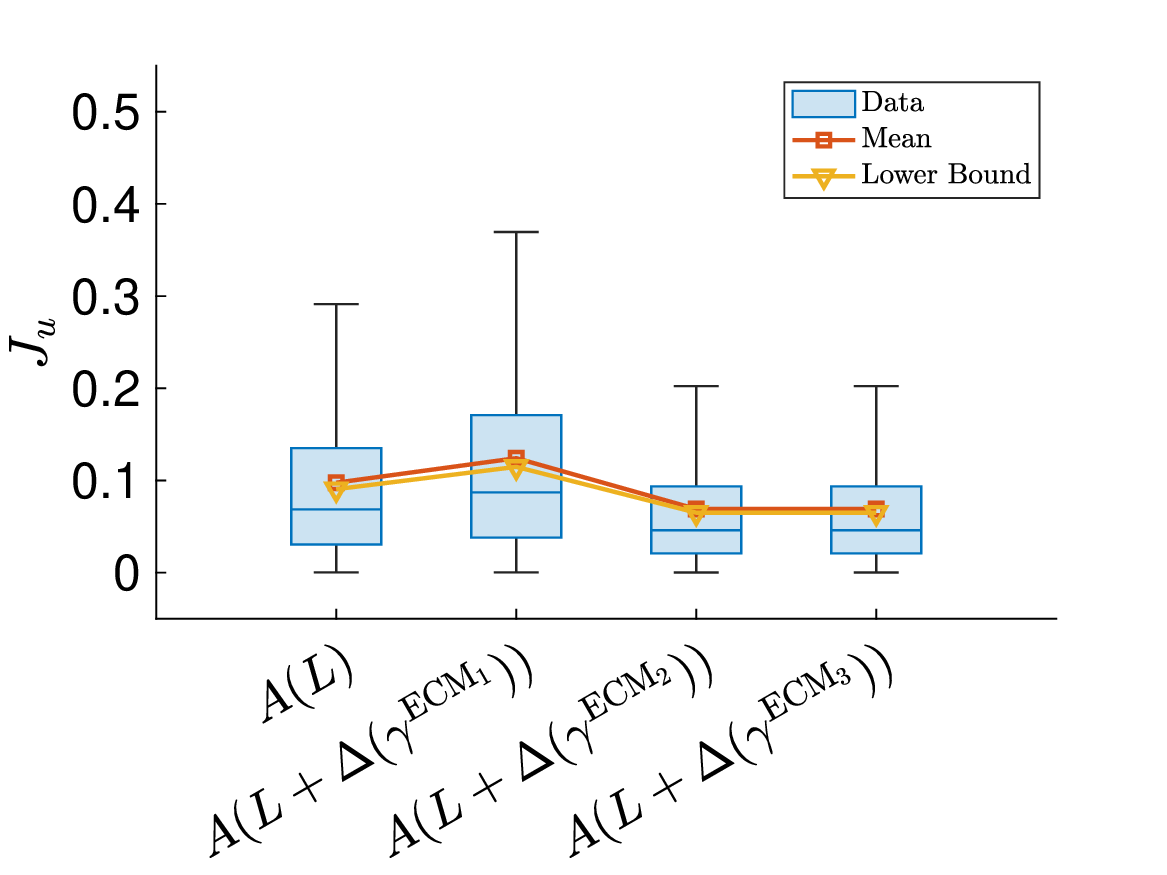}
    \includegraphics[scale = 0.3]{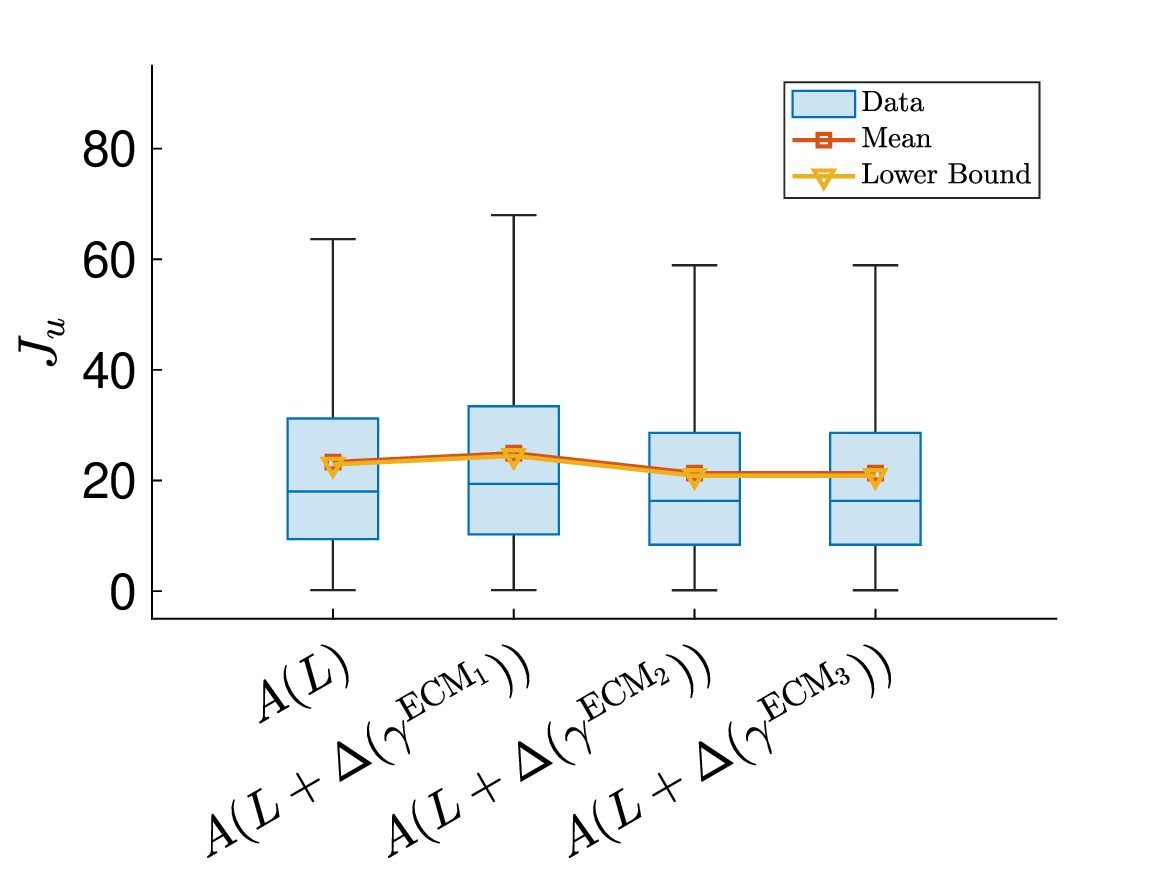}
    \includegraphics[scale = 0.3]{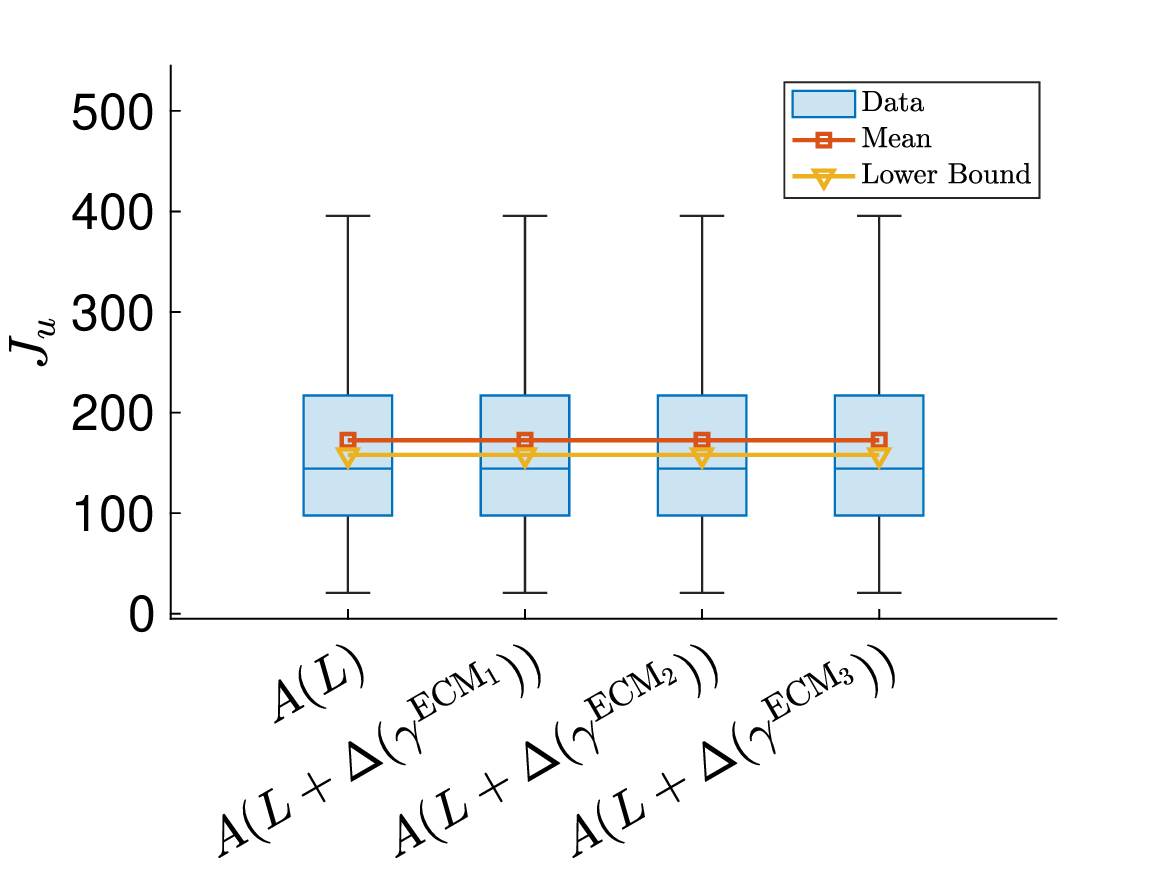}
    
    \includegraphics[scale = 0.3]{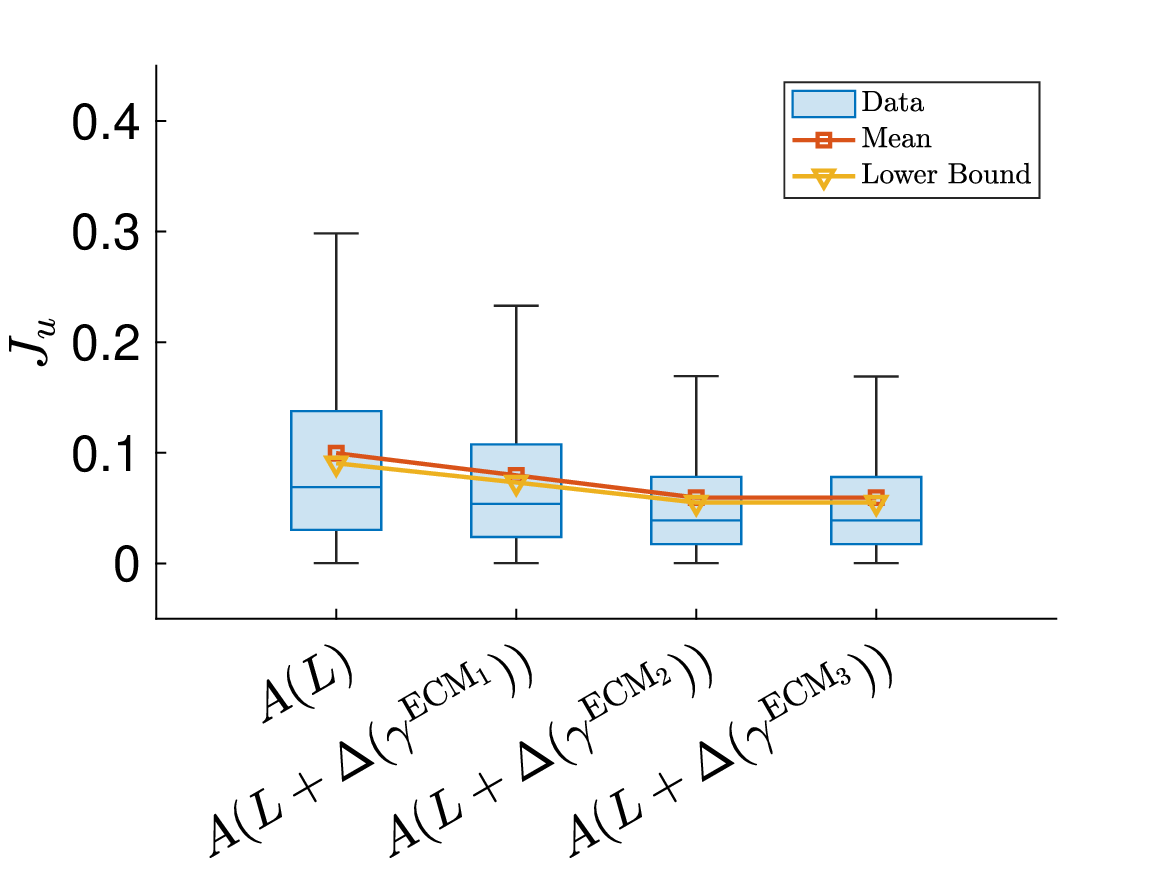}
    \includegraphics[scale = 0.3]{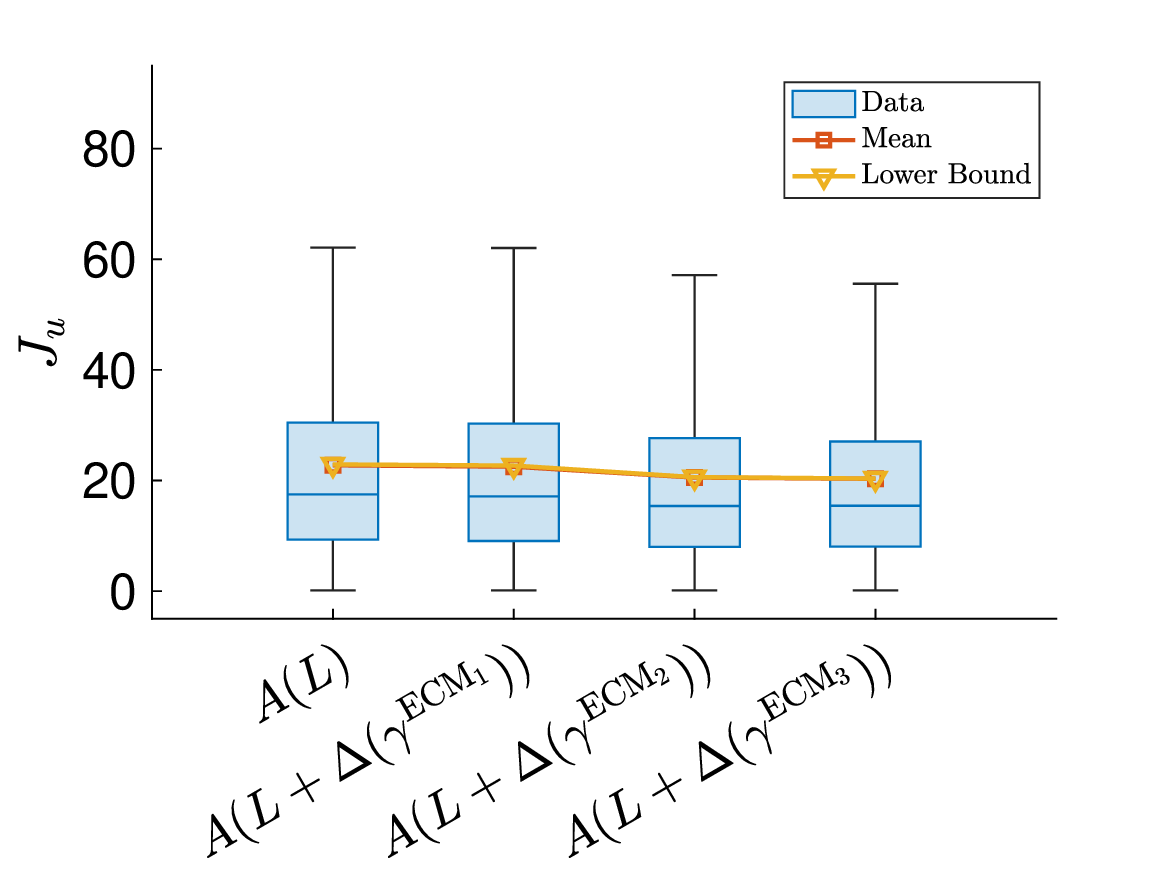}
    \includegraphics[scale = 0.3]{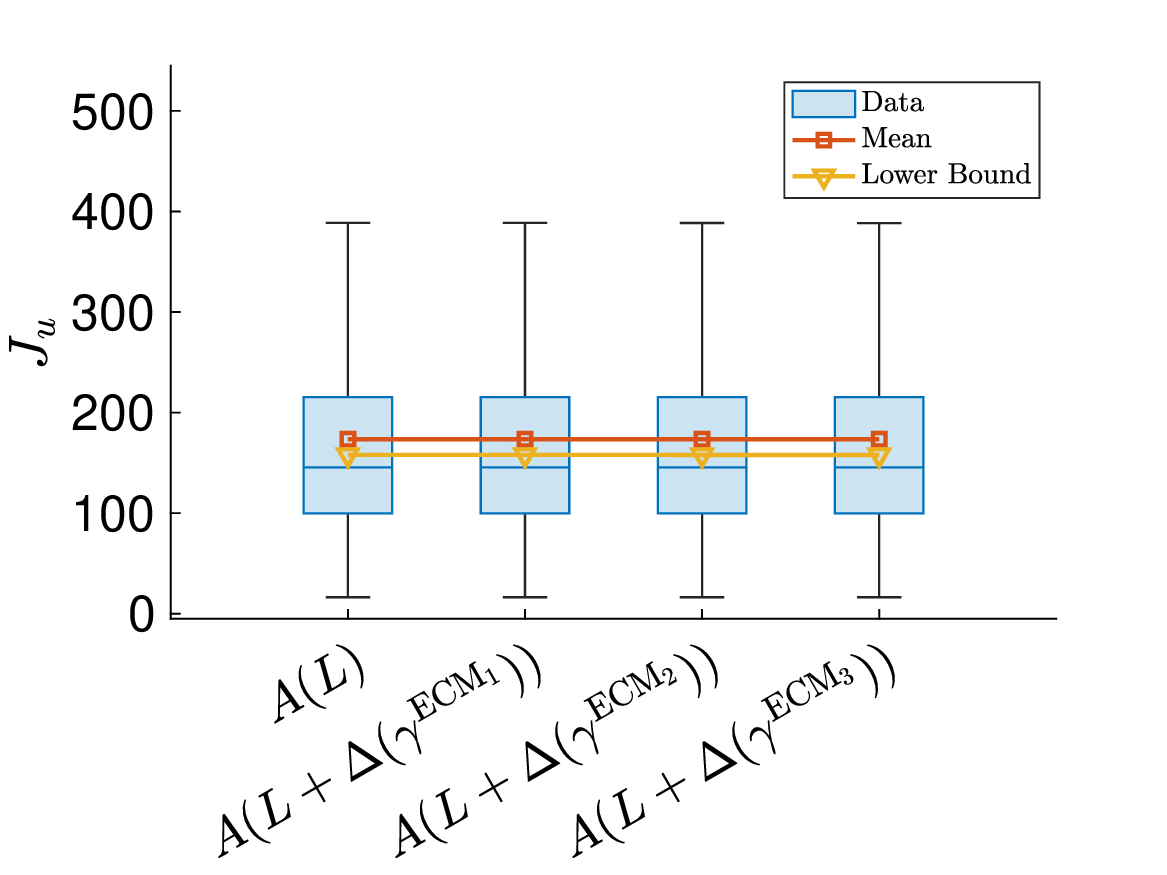}
    
    \caption{$J_u$ and the average minimum energy $\mathrm{tr}(W_c(t_f)^{-1})$ associated with the minimum energy input in \eqref{umin} for $t_f = -\frac{1}{\alpha(A(L))}$ and $t_f = \infty$ (lower bound) for these networks: IEEE $9$-bus, $s = 1$ (top-left) and $s=2$ (bottom-left), IEEE $14$-bus, $s = 1$ (top-middle) and $s=2$ (bottom-middle), IEEE $68$-bus, $s = 1$ (top-right) and $s=2$ (bottom-right).}
    \label{fig:enter-label-}
\end{figure*}

To conduct a comparison among the various choices of the Gramian-based performance metrics in \eqref{PerfoMet}, considering the various IEEE benchmarks and solving the continuous-time Lyapunov equation \eqref{ALCG} for $W_c(t_f)$ via the MATLAB built-in function $W_c(t_f) = \texttt{lyap}(A(\mathcal{L}),B B^\top - e^{At_f}B B^\top e^{A^\top t_f})$, we compute the corresponding minimum energy $J_u := x_0^\top W_c(t_f)^{-1} x_0$ for $10,000$ randomly generated $x_0$s (via MATLAB command \texttt{randn}) with $t_f = -\frac{1}{\alpha(A(L))}$. For the various choices of the Gramian-based performance metrics in \eqref{PerfoMet} and $s \in \{1,2\}$, Fig. \ref{fig:enter-label-} visualizes $J_u$ and the average (mean) minimum energy $\mathrm{tr}(W_c(t_f)^{-1})$ associated with the minimum energy control input in \eqref{umin} for $t_f = -\frac{1}{\alpha(A(L))}$ and $t_f = \infty$ (the lower bound). Remarkably, for any $x_0$, we have $x_0^\top W_c(t_f = \infty)^{-1} x_0 \le x_0^\top W_c(t_f = -\frac{1}{\alpha(A(L))})^{-1} x_0$ as it takes less energy to get somewhere more leisurely \cite{boyd2023controllability}. As depicted by Fig. \ref{fig:enter-label-}, for all the IEEE benchmarks, the third Gramian-based performance metric, i.e., the negated trace inverse performance metric achieves the best performance among the various choices of the Gramian-based performance metrics in \eqref{PerfoMet} which is not unexpected as the average (mean) minimum energy associated with the minimum energy control input in \eqref{umin} has the same form $\mathrm{tr}(W_c(t_f)^{-1})$. Also, we observe that the log-det-based solution similarly improves the average (mean) minimum energy $\mathrm{tr}(W_c(t_f)^{-1})$ associated with the minimum energy control input in \eqref{umin} while the trace-based solution may or may not improve such average (mean) minimum energy $\mathrm{tr}(W_c(t_f)^{-1})$. In summary, we highlight that the second and third Gramian-based performance metrics attain performance improvements compared to the nominal case $A(L)$ while the first Gramian-based performance metric may or may not attain a performance improvement aligned with its weakest functionality observed in previous observations. Furthermore, the negated trace inverse performance metric achieves the best performance among the various choices of the Gramian-based performance metrics in \eqref{PerfoMet} in terms of the average (mean) minimum energy $\mathrm{tr}(W_c(t_f)^{-1})$ associated with the minimum energy control input in \eqref{umin}.

\subsection{Parametric dependency of performance improvement}

\begin{figure}[t]
    \centering
    \includegraphics[scale = 0.22]{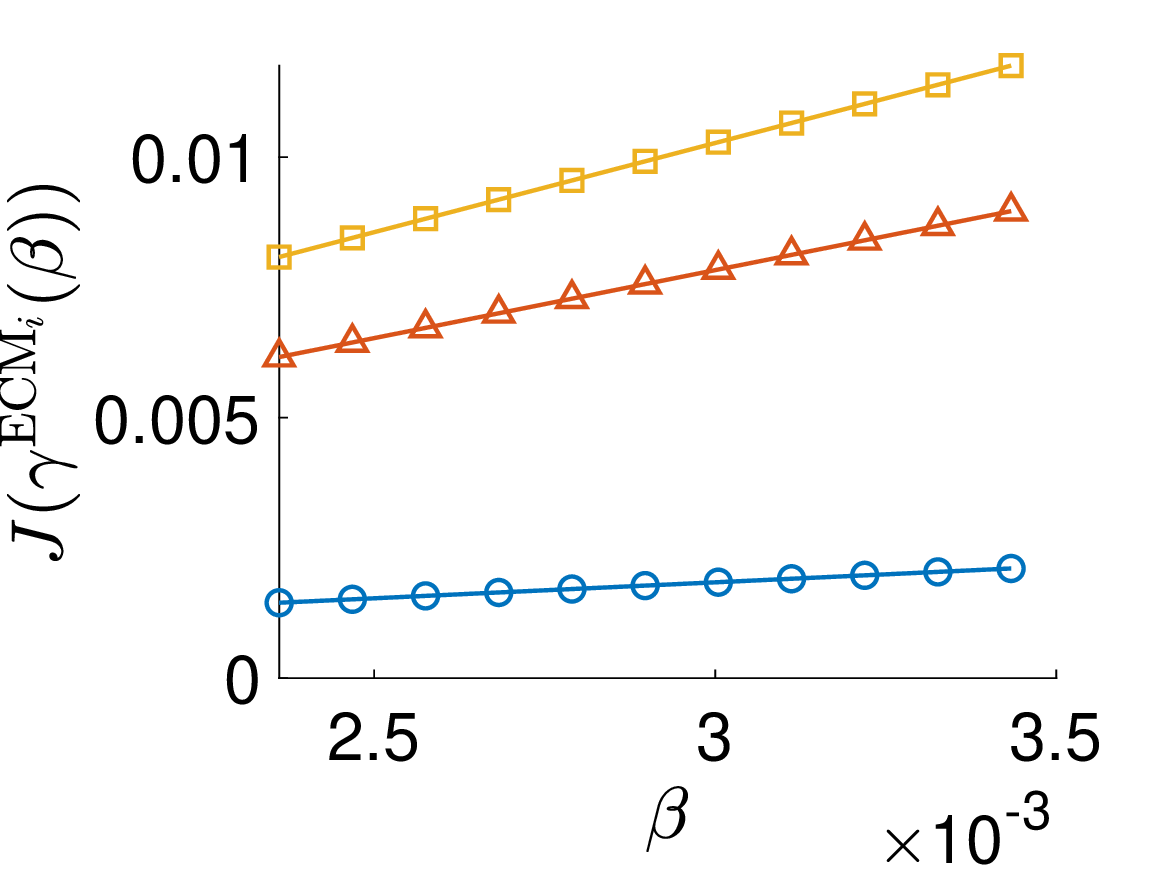}
    \includegraphics[scale = 0.22]{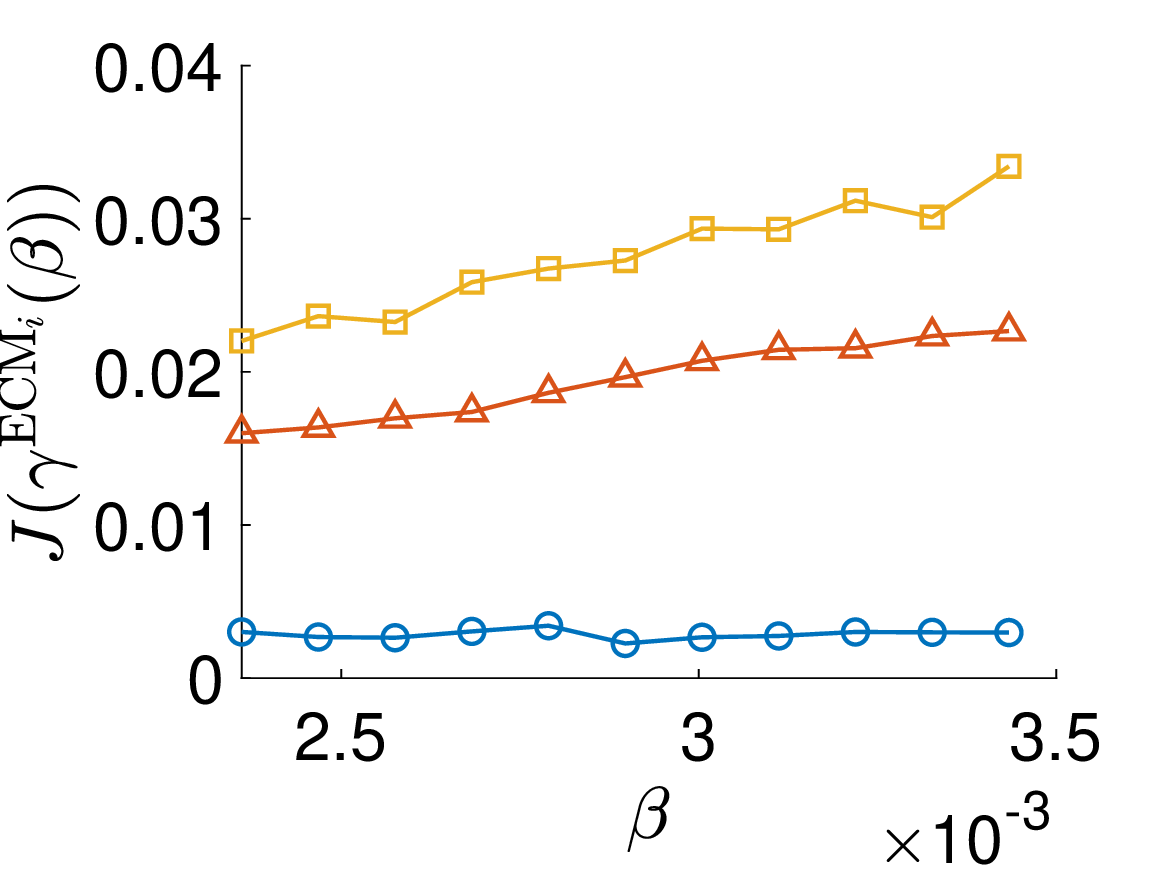}
    \caption{Relationship between the performance improvement $J(\gamma^{\mathrm{ECM}_i}(\beta))~(\%)$ and the budget/energy parameter $\beta$. IEEE $68$-bus benchmark, $i \in \{1,2,3\}$, $s = 1$ (left) and $s=15$ (right). $i = 1$: blue circle, $i=2$: red triangle, and $i = 3$: yellow square.}
    \label{fig:enter-label-2}
\end{figure}

Let us empirically study the relationship between the performance improvement $J(\gamma^{\mathrm{ECM}_i}(\beta))$ and the budget/energy parameter $\beta$. Running Procedure \ref{pro:two} for the equidistant values of $\beta$, we get Fig. \ref{fig:enter-label-2} that illustrates the relationship between the performance improvement $J(\gamma^{\mathrm{ECM}_i}(\beta))$ and the budget/energy parameter $\beta$ for $s = 1$ (left) and $s=15$ (right), considering the IEEE $68$-bus benchmark and $i \in \{1,2,3\}$. We observe that the larger the budget/energy upper-bound we consider, the higher performance improvement we achieve. Such observation is not unexpected as in optimization problem \eqref{OPn}, increasing the value of $\beta$, the feasibility set expands according to \eqref{OPCn}. Also, note that in the case of the IEEE $68$-bus benchmark, similar to the previous observations, the first Gramian-based performance metric demonstrates the weakest functionality, unlike the other Gramian-based performance metrics that significantly demonstrate an increasing trend in Fig. \ref{fig:enter-label-2} (right).

\begin{figure}[t]
    \centering
    \includegraphics[scale = 0.22]{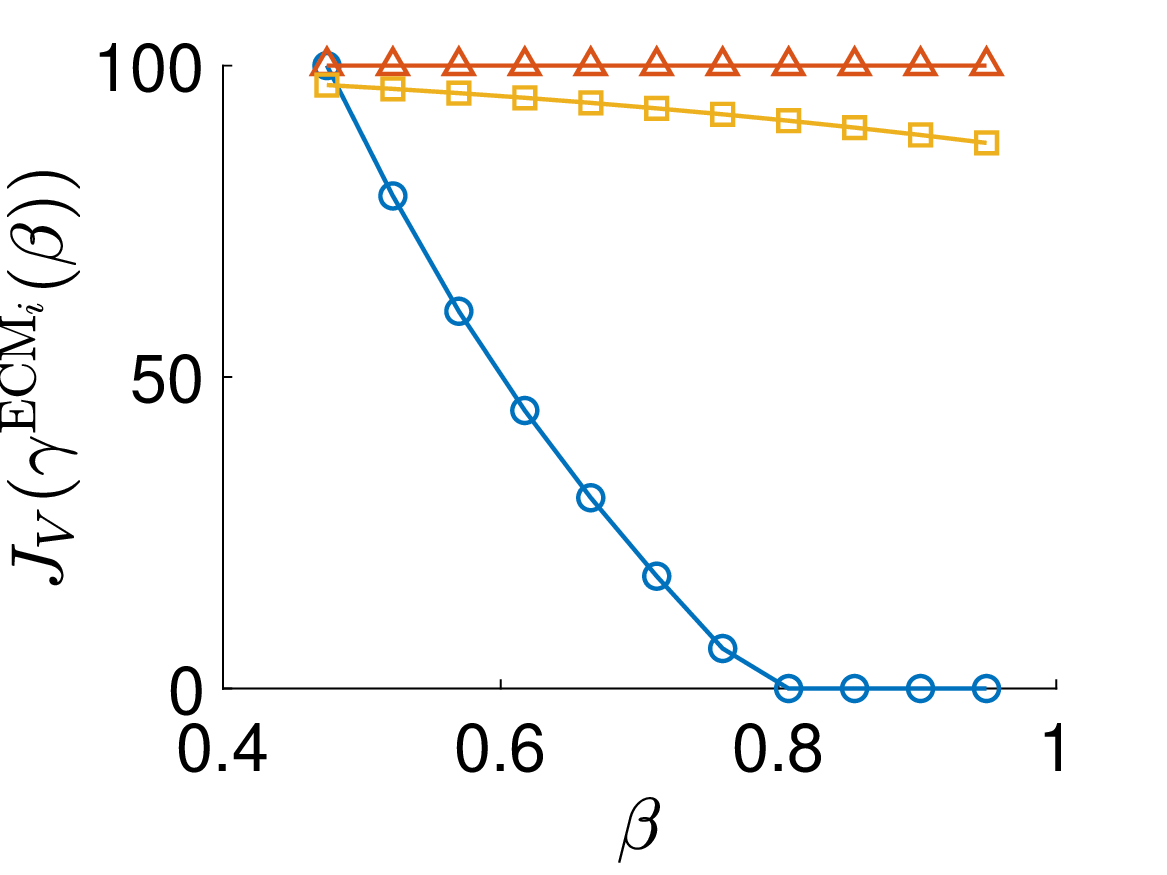}
    \includegraphics[scale = 0.22]{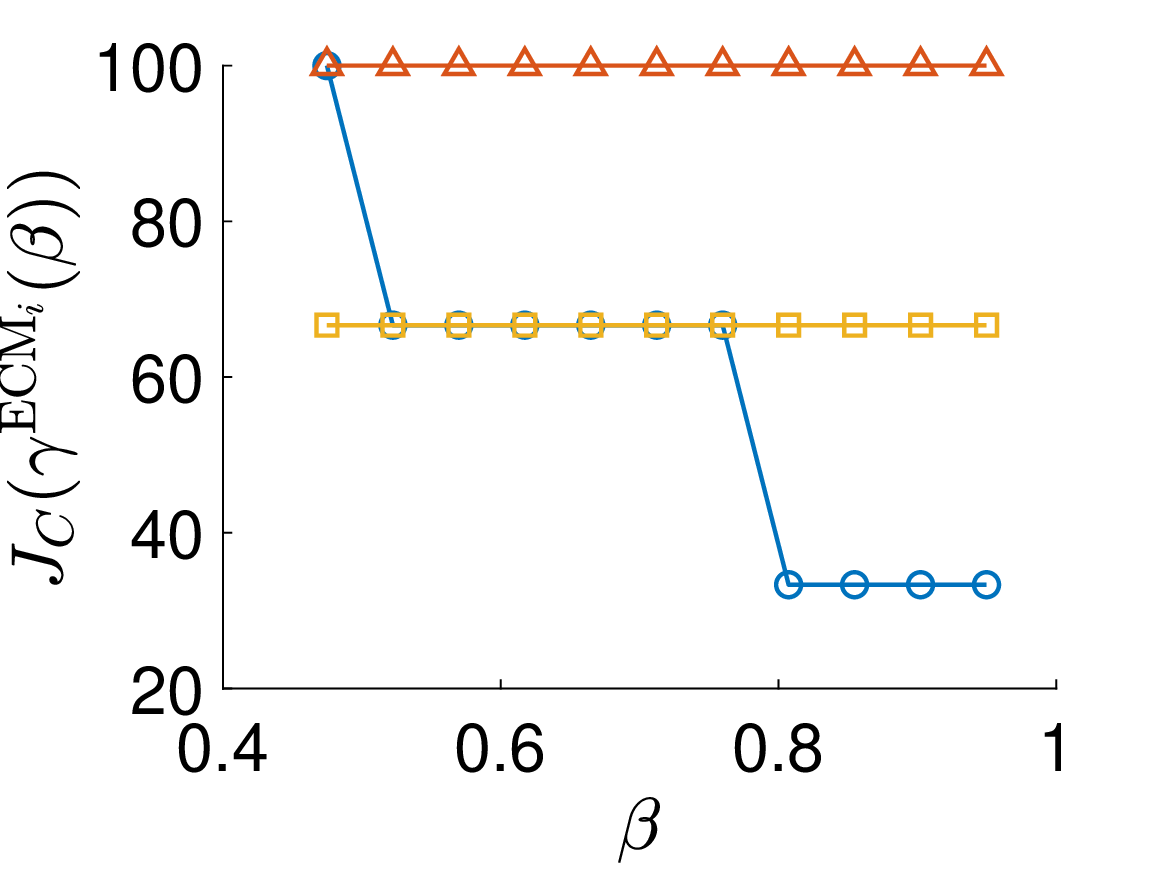}
    
    \caption{Relationship between the near-optimality performance measures $J_V(\gamma^{\mathrm{ECM}_i}(\beta))~(\%)$ and $J_C(\gamma^{\mathrm{ECM}_i}(\beta))~(\%)$ and the budget/energy parameter $\beta$ for the IEEE $9$-bus benchmark with $\beta \le \min\{g_{ji}: (i,j) \in \mathcal{E}\}$, $s = 1$ and $i \in \{1,2,3\}$. $i = 1$: blue circle, $i=2$: red triangle, and $i = 3$: yellow square.}
    \label{fig:enter-label-3}
\end{figure}

As shown in Tabs. \ref{tab:my_label1} and \ref{tab:my_label2} for the IEEE $9$-bus benchmark with { $\beta = 1$ ($\beta > \min\{g_{ji}: (i,j) \in \mathcal{E} \}$),} $s = 1$, and $g(W_c) = \mathrm{tr}(W_c)$, the near-optimality performances are less than $100~\%$. Fig. \ref{fig:enter-label-3} depicts the relationship between the near-optimality performance measures $J_V(\gamma^{\mathrm{ECM}_i}(\beta))$ and $J_C(\gamma^{\mathrm{ECM}_i}(\beta))$ and the budget/energy parameter $\beta$ for the IEEE $9$-bus benchmark with $\beta \le \min\{g_{ji}: (i,j) \in \mathcal{E}\}$, $s = 1$, and $i \in \{1,2,3\}$. As depicted by Fig. \ref{fig:enter-label-3}, for sufficiently small values of the budget/energy upper-bound imposed by $\beta \le \min\{g_{ji}: (i,j) \in \mathcal{E}\}$, the smaller the budget/energy upper-bound we consider, the better near-optimality performance we obtain. Such observation can be justified due to the derivative-based nature of the performance sensitivity information captured by the edge centrality notion. In other terms, the edge centrality matrix can detect the most influential edges perfectly for sufficiently small edge perturbations (in the norm sense). For large edge perturbations (in the norm sense), the edge centrality matrix can potentially become impractical. Then, the ECM-based edge modification method achieves its best performance in the case of low-budget/energy upper-bound scenarios. 

\subsection{Damping performance} \label{DampPerf}

\begin{figure*}[t]
    \centering
    \includegraphics[scale = 0.3]{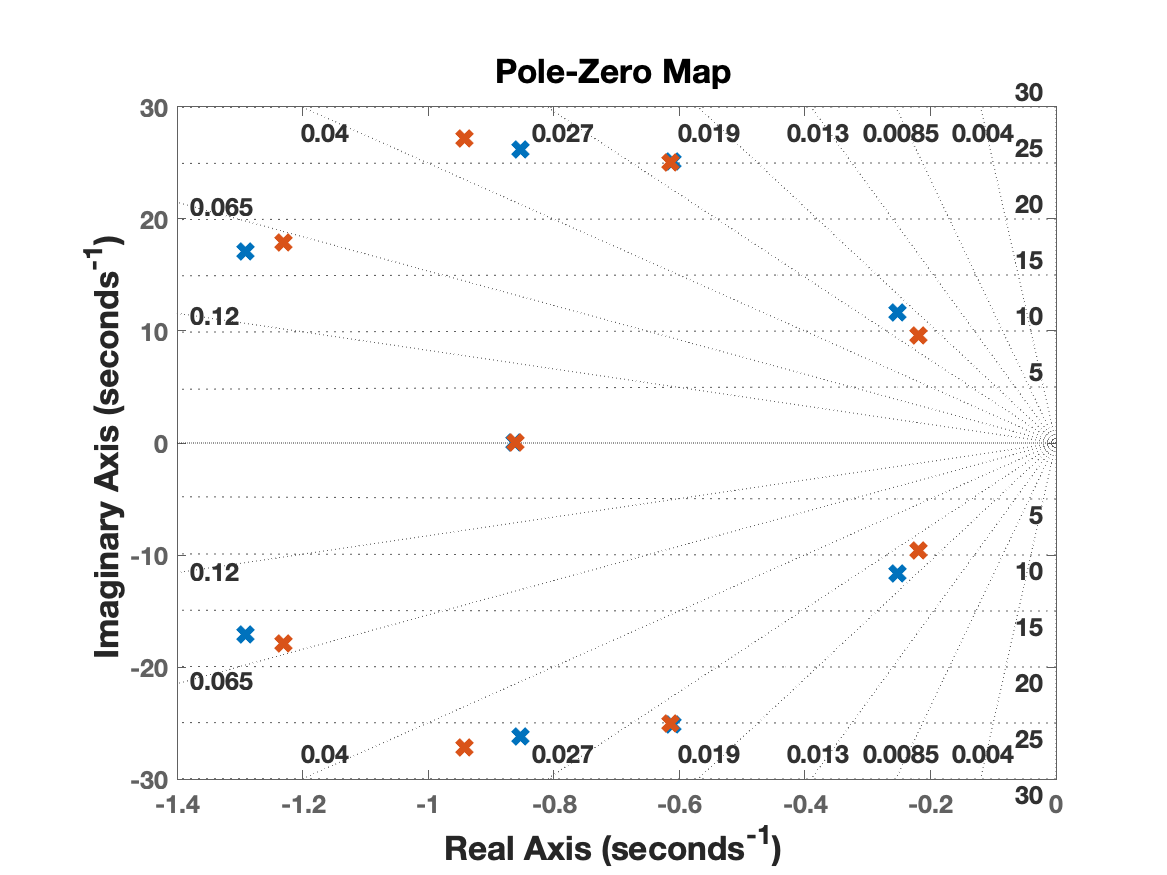}
    \includegraphics[scale = 0.3]{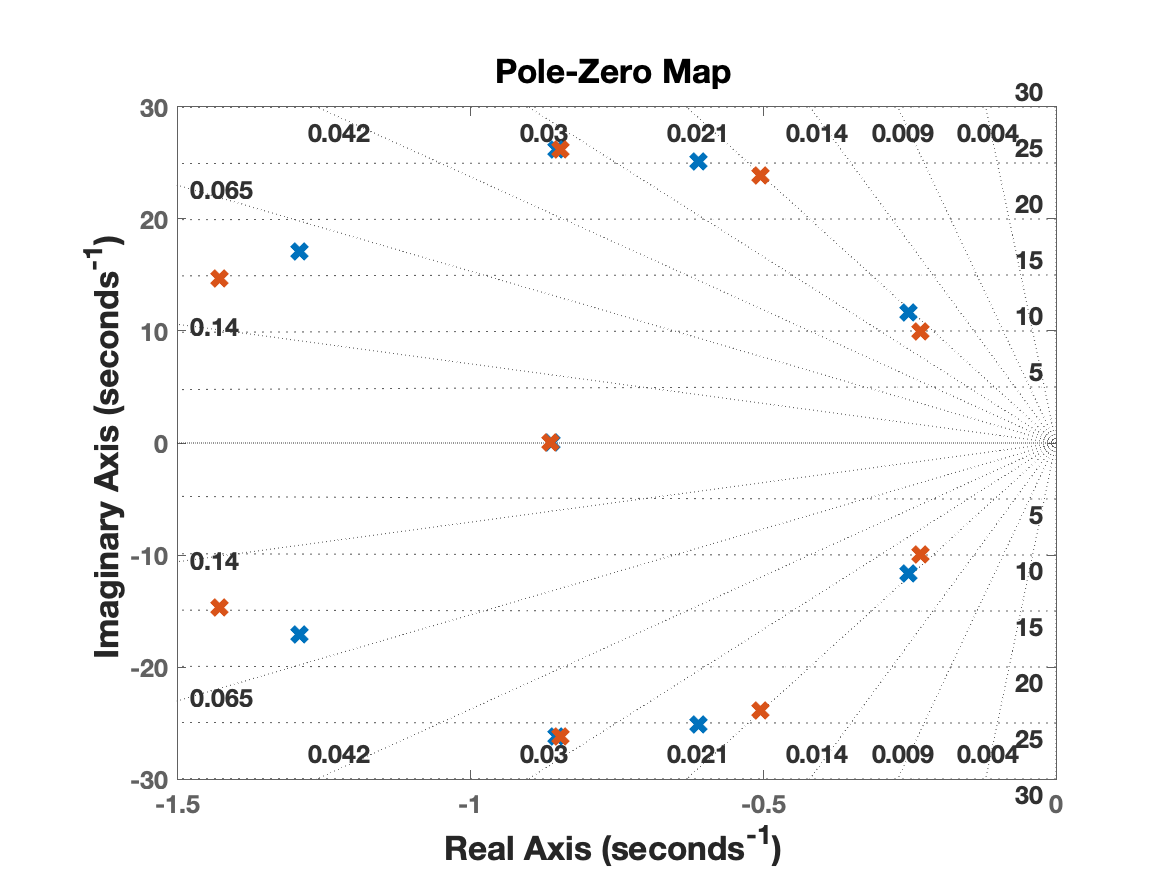}
    \includegraphics[scale = 0.3]{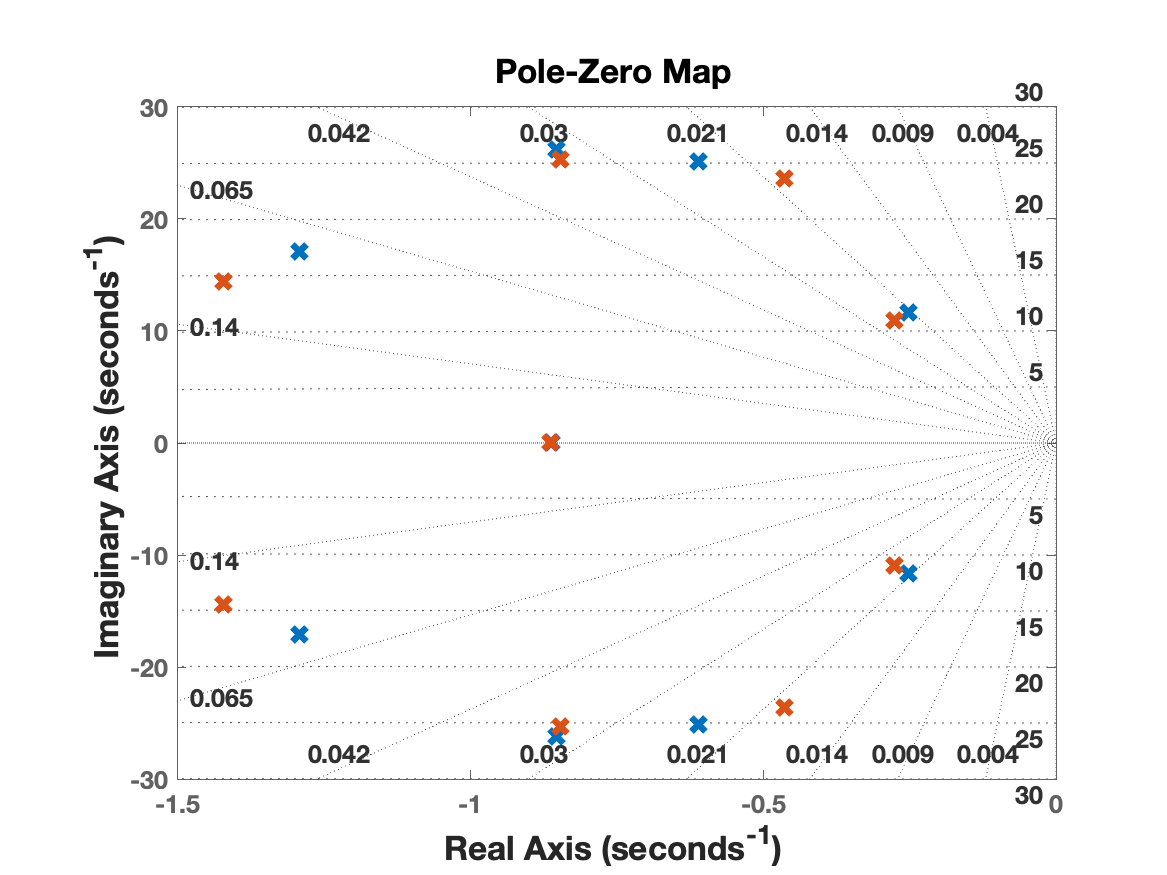}
    \caption{Pole-zero maps with damping lines corresponding to the $A(L)$ and $A(L + \Delta (\gamma^{\mathrm{ECM}_i}))$ for $i \in \{1,2,3\}$ and $s = 2$. IEEE $14$-bus, from $i = 1$ (left) to $i = 3$ (right). The crosses in blue and red correspond to the poles of $A(L)$ and $A(L + \Delta (\gamma^{\mathrm{ECM}_i}))$.}
    \label{fig:enter-label-5}
\end{figure*}

In this section, considering the IEEE $14$-bus benchmark, we visualize the pole-zero maps with damping lines (via MATLAB command \texttt{pzmap}) corresponding to the $A(L)$ and $A(L + \Delta (\gamma^{\mathrm{ECM}_i}))$ for $i \in \{1,2,3\}$ and $s = 2$ in Fig. \ref{fig:enter-label-5}. For a given non-zero pole $p$, its damping ratio ($\%$) is defined as
{\begin{align} \label{DR}
    \zeta(p) &:= 100 \times {-\Re(p)}/{\sqrt{\Re(p)^2 + \Im(p)^2}}.  
\end{align}}The damping ratio $\zeta(p)$ ($\%$) defined by \eqref{DR} can geometrically be interpreted as the cosine of the acute angle between the damping line and the real axis. Thus, the lower the angle value we have, the higher the damping ratio we get. As depicted by Fig. \ref{fig:enter-label-5}, for the IEEE $14$-bus benchmark, the slow modes, i.e., the modes close to the imaginary axis, have been dampened effectively. The edge modifications associated with the Gramian-based performance metrics $\mathrm{tr}(W_c(t_f))$, $\log \det(W_c(t_f))$, and $-\mathrm{tr}(W_c(t_f)^{-1})$ attain $5.22~\%$, $6.83~\%$, and $16.95~\%$ damping ratio improvements (for the slow modes) compared to the nominal network, respectively where the corresponding damping ratios $\zeta$ ($\%$) for $A(L)$ and $A(L + \Delta (\gamma^{\mathrm{ECM}_i}))$ are $2.17~\%$, $2.28~\%$, $2.31~\%$, and $2.53~\%$, respectively. We observe that the third Gramian-based performance metric significantly outperforms the other Gramian-based performance metrics in terms of damping ratio improvement (for the slow modes). However, to systematically enhance the damping coefficients, incorporating a regularization term in the optimization problems seems necessary.

\subsection{Reachability analysis}

\begin{figure*}
    \centering
    \includegraphics[scale = 0.3]{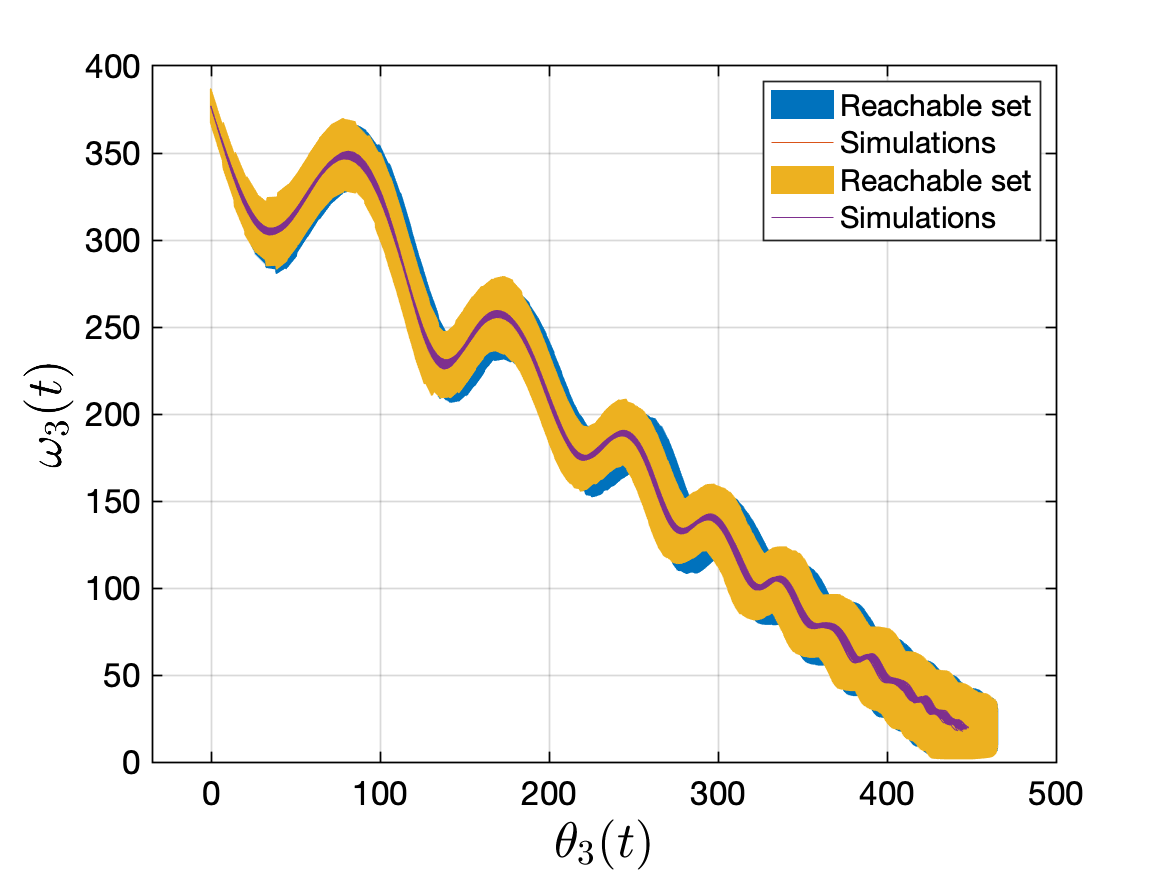}
    \includegraphics[scale = 0.3]{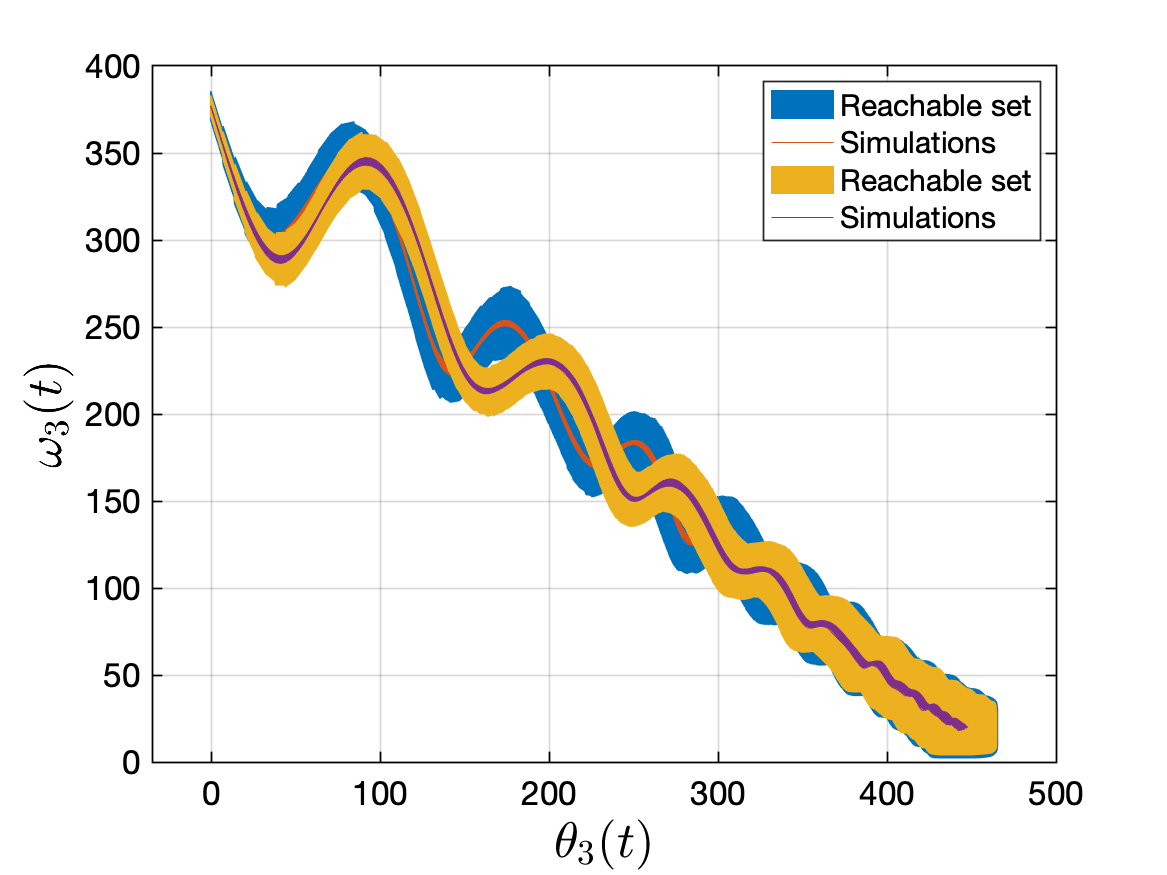}
    \includegraphics[scale = 0.3]{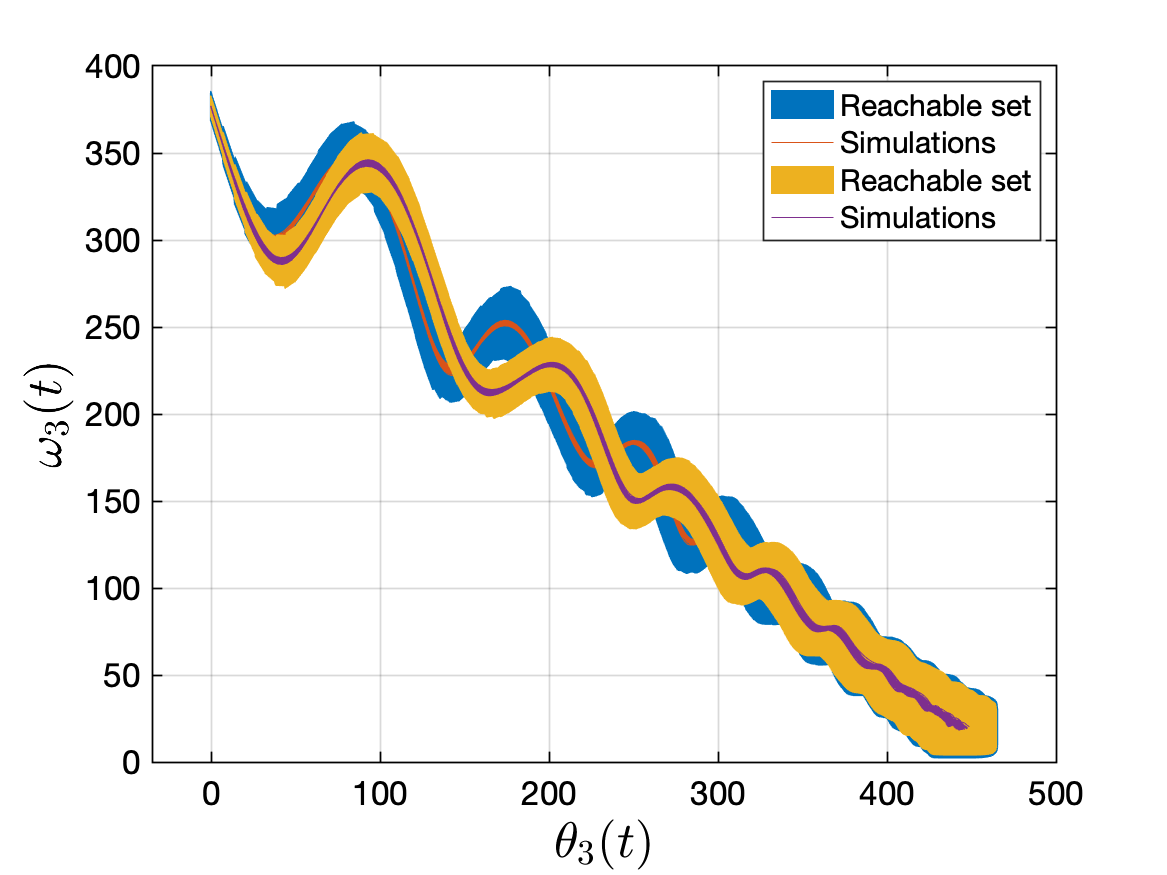}
    \caption{Reachable sets corresponding to the Generator $3$ (its rotor angle $\theta_3(t)$ and rotor frequency $\omega_3(t)$) associated with the original network $\bar{A}(L)$ and the modified network $\bar{A}(L + \Delta (\gamma^{\mathrm{ECM}_i}))$ for $i \in \{1,2,3\}$ and $s = 2$. IEEE $14$-bus, from $i = 1$ (Left) to $i = 3$ (Right). The blue reachable set and red simulations commonly refer to the case of the original network $\bar{A}(L)$ and the yellow reachable set and violet simulations refer to the corresponding modified network $\bar{A}(L + \Delta (\gamma^{\mathrm{ECM}_i}))$ for $i \in \{1,2,3\}$.}
    \label{Fig:Gen3}
\end{figure*}

In this section, considering the IEEE $14$-bus benchmark, we visualize the reachable sets associated with the $\bar{A}(L)$ and $\bar{A}(L + \Delta (\gamma^{\mathrm{ECM}_i}))$ for $i \in \{1,2,3\}$ and $s = 2$. Starting from the zonotope-like initial sets around
{\begin{align*}
    x_0 &= \begin{bmatrix}
    \theta_0^\top & \omega_0^\top
\end{bmatrix}^\top,~\omega_0 = 120 \pi \mathbf{1}_N,\\
    \theta_0 &= \begin{bmatrix}
    0.1432 & 0.0056 & -0.2221 & -0.1800 & -0.1531  
\end{bmatrix}^\top,
\end{align*}}with $\pm 0.1 \mathbf{1}_{2N}$ as initial state variation bounds and considering the randomly generated bounded inputs $u(t)$ around 
{\begin{align*}
    u_0 &= \begin{bmatrix}
    2.3239 & 0.4000 & -0.0000 & 0.0000 & -0.0000
\end{bmatrix}^\top,
\end{align*}}with $\pm \texttt{rand(N,1)}$ as input variation bounds, by utilizing the reachability analysis toolbox continuous reachability analyzer (CORA) developed by \cite{althoff2024cora}, we visualize the reachable sets corresponding to the Generator $3$ (its rotor angle $\theta_3(t)$ and rotor frequency $\omega_3(t)$) in the original network $\bar{A}(L)$ and the modified network $\bar{A}(L + \Delta (\gamma^{\mathrm{ECM}_i}))$ for $i \in \{1,2,3\}$ in Fig. \ref{Fig:Gen3}. Note that reachable sets visualized in Fig. \ref{Fig:Gen3} are over-approximations of the randomly created simulations (created from $100$ random samples from the initial set around $x_0$). As depicted by Fig. \ref{Fig:Gen3}, for the IEEE $14$-bus benchmark, the modified network associated with the third Gramian-based performance metric attains the best state trajectories improvement (among three performance metrics) vertically and horizontally towards the final reachable sets compared to the original network. Such an observation is aligned with the satisfactory results of the third Gramian-based performance metric in terms of damping performance studied in Section \ref{DampPerf}. We also highlight that the reachability analysis in this section has been conducted via edge modifications and a more realistic scenario can potentially be done by altering the admittance matrix. Such a more realistic scenario is left as a future direction.

\section{Concluding Remarks and Future Directions} \label{Con}
Here, we provide the answers \textit{A1}--\textit{A6} to the corresponding research questions \textit{Q1}--\textit{Q6} listed in Section \ref{secNS} as follows: 

{\begin{itemize}
    \item \textit{A1}: \textit{Yes.} We first quantitatively assess the edge perturbation impact on simple linear swing dynamics of the synchronous generators by utilizing a recently introduced control-theoretic powerful notion of edge centrality matrix (ECM). Then, we efficiently design the edge modification vector by solving an optimization problem for a near-optimal edge modification to strengthen the controllability of the power network.

    \item \textit{A2}: Various numerical experiments on the IEEE power network benchmarks reveal that the proposed ECM-based edge modification method achieves an acceptable level of near-optimality compared to the best-case and worst-case scenarios obtained from the brute-force exhaustive combinatorial search in the case of computationally feasible circumstances (e.g., the low-cardinality edge modification scenario).

    \item \textit{A3}: Since identifying the optimal edge modification vector requires a brute-force exhaustive combinatorial search (the NP-hardness), the ECM-based approach significantly reduces the computational burden. Furthermore, for the high-cardinality edge modification scenario, the ECM-based edge modification method proposes a solution that outperforms the random-based solution.

\item \textit{A4}: \textit{No.} The simulations corroborate that various Gramian-based performance metrics showcase different behavior. In particular, the negated trace inverse performance metric achieves the best performance among the various choices of the Gramian-based performance metrics in terms of the average (mean) minimum energy associated with the minimum energy control input.

\item \textit{A5}: Based on empirical validations, we observe that a higher budget/energy upper-bound leads to better performance improvement that makes sense. Also, we observe that the ECM-based edge modification method achieves its best performance in the case of low-budget/energy upper-bound scenarios due to the derivative-based nature of the performance sensitivity information captured by the edge centrality notion.

\item \textit{A6}: Several numerical experiments on the IEEE power network benchmarks showcase that the proposed ECM-based method outperforms a static graph-theoretic edge centrality-based alternative, namely nearest neighbor edge centrality (NNEC) \cite{brohl2022straightforward}, in identifying the network's most influential edges.
    
\end{itemize}}

\noindent{\it Limitations:} Here, we list some limitations of this study:
\begin{itemize}
\item No theoretical near-optimality guarantees are presented for the ECM-based edge modification method. Only in the case of computationally feasible circumstances (e.g., the low-cardinality edge modification scenario), one can empirically assess the near-optimality performance of the proposed method.  

\item The near-optimality performance of the ECM-based edge modification method can potentially deteriorate by increasing the budget/energy upper-bound because, for large edge perturbations (in the norm sense), the edge centrality matrix can potentially become impractical.

\item Although the linear power dynamics utilized in this paper provide us with a convenient framework to propose edge modification designs to improve the power system controllability, it is simplistic and still far away from the more representative (realistic) nonlinear differential-algebraic equations (NDAEs)-modeled power systems.

\end{itemize}

\noindent{\it Future directions:} A potential future direction could be improving the scalability and near-optimality via proposing tailored optimization techniques. Another pertinent future direction could be generalizing the results of this paper (the systems with linear dynamics) to those with nonlinear dynamics as they would be more representative, being suitable for realistic scenarios. 

\bibliographystyle{IEEEtran}
\bibliography{References}

\begin{IEEEbiography}[{\includegraphics[width=1in,height=1.25in,clip,keepaspectratio]{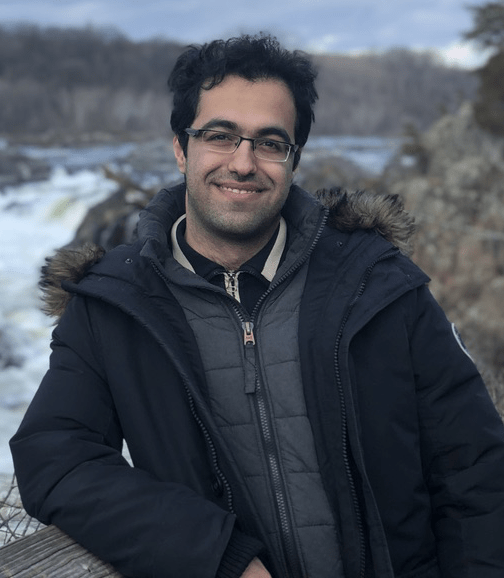}
 }] {MirSaleh Bahavarnia} (Member, IEEE) received the B.Sc. degree in electrical engineering (control) and a certificate of the minor program in mathematics from the Sharif University of Technology, Tehran, Tehran, Iran, in 2013 and the Ph.D. degree in mechanical engineering (control) from the Lehigh University, Bethlehem, PA, USA, in 2018.

From 2018 to 2020, he was a Postdoctoral Research Associate with the Department of Electrical and Computer Engineering and the Institute for Systems Research (ISR), University of Maryland, College Park, MD, USA. Since 2022, he has been a Postdoctoral Research Scholar with the Department of Civil and Environmental Engineering, Vanderbilt University, Nashville, TN, USA. His research interests include distributed control, feedback control, power systems control, process control, robust control, and traffic control.
\end{IEEEbiography}

\begin{IEEEbiography}[{\includegraphics[width=1in,height=1.25in,clip,keepaspectratio]{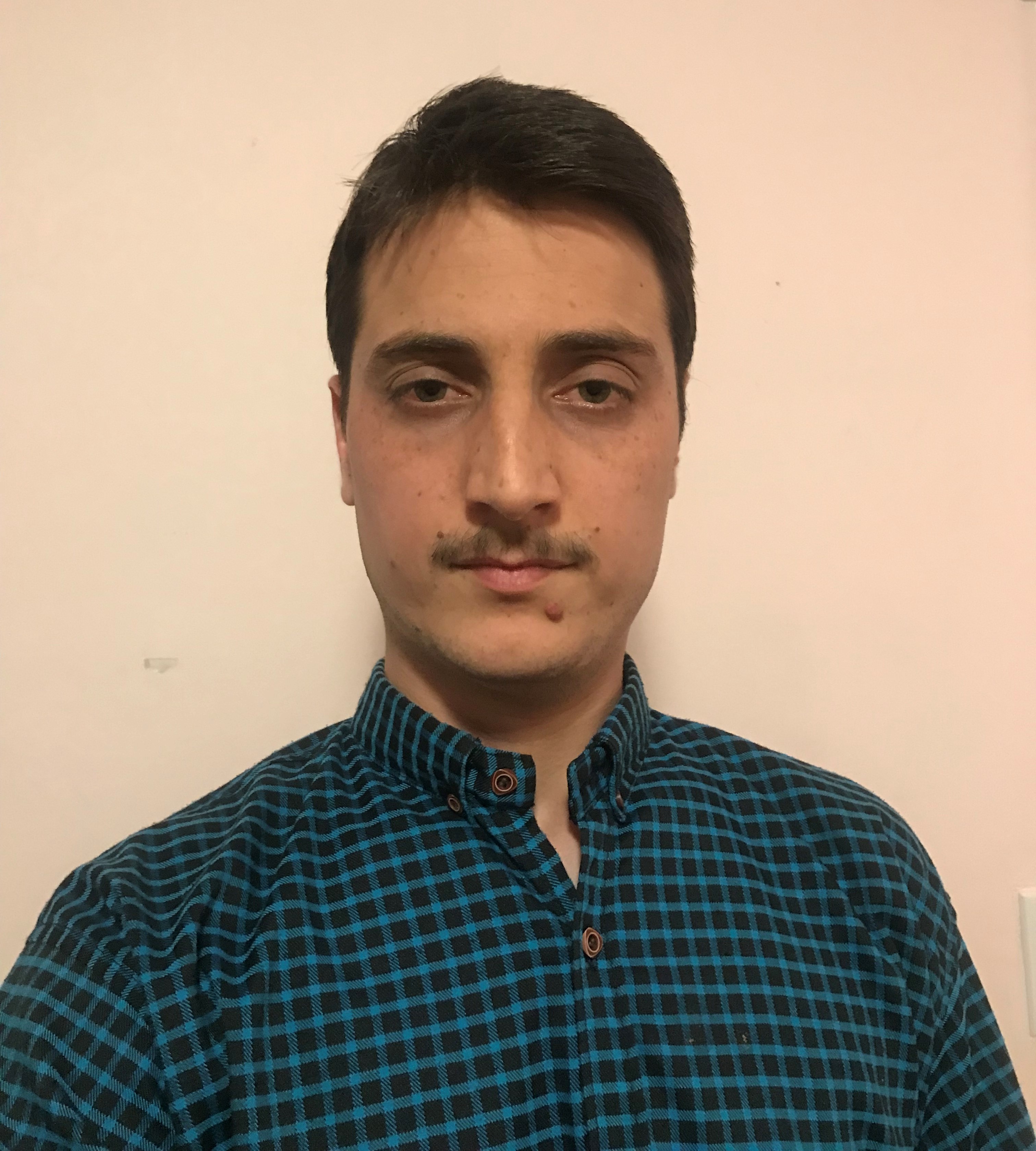}}]
	{Muhammad Nadeem} (Member, IEEE) was born in Nilore, Islamabad, Pakistan. He received the B.E. and M.S. degrees in electrical engineering (power) from Air University and National University of Science and Technology (NUST), Islamabad, Pakistan, in 2017 and 2020, respectively. He received the Ph.D. degree in Civil Engineering from Vanderbilt University, Nashville, TN, USA, in 2025. His research interests include dynamic state estimation and control theory for renewables-heavy power systems.
\end{IEEEbiography}

\begin{IEEEbiography}[{\includegraphics[width=1in,height=1.25in,clip,keepaspectratio]{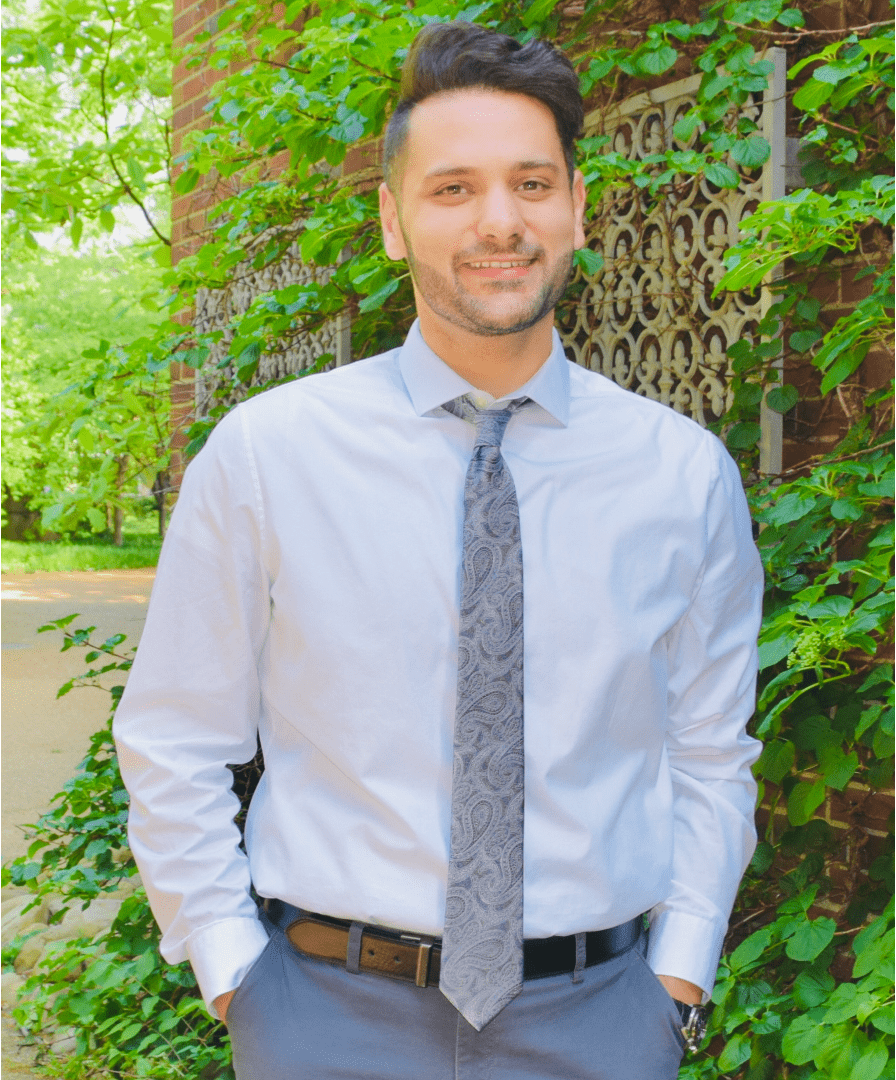}
 }] {Ahmad F. Taha} (Member, IEEE) received the B.E. degree in electrical and computer engineering from the American University of Beirut, Beirut, Lebanon, in 2011, and the Ph.D. degree in electrical and computer engineering from Purdue University, West Lafayette, IN, USA, in 2015. 
 
Before joining Vanderbilt University, Nashville, TN, USA, he was an Assistant Professor with the ECE Department, at the University of Texas, San Antonio. He is an Associate Professor with the Department of Civil and Environmental Engineering, at Vanderbilt University. He has a secondary appointment in the ECE Department. His research interests include understanding how complex cyber-physical and urban infrastructures operate, behave, and occasionally misbehave, and optimization, control, monitoring, and security of infrastructure with power, water, and transportation systems applications.

Dr. Taha is an Associate Editor for IEEE Transactions on Control of Network Systems.

\end{IEEEbiography}

\end{document}